\documentclass[twocolumn]{svjour3}  

\usepackage{xspace}
\usepackage{tikz,pifont,pgfplots}
\usepackage{xcolor, colortbl}
\usepackage{makecell,multirow,diagbox}
\usepackage{hyperref}
\usepackage[many]{tcolorbox}
\usepackage{graphicx}%
\usepackage{adjustbox}
\usepackage{amsmath,amssymb,amsfonts}%
\usepackage{mathrsfs}%
\usepackage[title]{appendix}%
\usepackage{xcolor}%
\usepackage{textcomp}%
\usepackage{manyfoot}%
\usepackage{booktabs}%
\usepackage{algorithm}%
\usepackage{algorithmicx}%
\usepackage{algpseudocode}%
\usepackage{listings}%
\usepackage{soul}
\usepackage{subcaption}
\usetikzlibrary{positioning}
\pgfplotsset{compat=1.18}

\newcommand{\rqfirst}{\textbf{RQ$_1$}: \emph{How do state-of-the-art toolkits allow users to explicitly specify customized fairness definitions?}} 
\newcommand{\rqsecond}{\textbf{RQ$_2$}: \emph{Can \tool support the whole fairness analysis process, including the automated assessment phase?}}
\newcommand{\rqthree}{\textbf{RQ$_3$}: \emph{Is \tool able to overcome the limitations of current MDE-based approaches for fairness assessment?}}

\newcommand*{\ie}{i.e.,\@\xspace}
\newcommand*{\eg}{e.g.,\@\xspace}

\newcommand*{\tool}{MODNESS\@\xspace}

\newcommand*{\etal}{\emph{et~al.}\@\xspace}

\newboolean{showrevision}
\setboolean{showrevision}{false}
\ifthenelse{\boolean{showrevision}}
{
	\newcommand\revision[1]{\textcolor{blue}{#1}}
}
{
	\newcommand\revision[1]{{#1}}
}

\newcommand\rev[1]{\textcolor{black}{#1}}

\definecolor{verylightgray}{gray}{0.92}
\definecolor{ao(english)}{rgb}{0.0, 0.5, 0.0}

\definecolor{deepblue}{rgb}{0,0,0.5}
\definecolor{deepred}{rgb}{0.6,0,0}
\definecolor{deepgreen}{rgb}{0,0.5,0}
\definecolor{shadecolor}{gray}{0.9}

\colorlet{shadecolor}{verylightgray}
\colorlet{framecolor}{black}

\newcommand\untick{\ding{55}}
\newcommand\tick{\ding{52}}

\newcommand*\circled[1]{\tikz[baseline=(char.base)]{\color{black} 
    \node[shape=circle,draw=cyan,fill=black!10!white,inner sep=.3pt] (char) {{{\texttt\textbf #1}}};}}

\newcommand*\circledb[1]{\tikz[baseline=(char.base)]{\color{black} 
    \node[shape=circle,draw=cyan,fill=blue!30!white,inner sep=.3pt] (char) {{{\texttt\textbf #1}}};}}

\newcommand*\circledw[1]{\tikz[baseline=(char.base)]{\color{black} 
    \node[shape=circle,draw=black,inner sep=.3pt] (char) {{{\texttt\textbf #1}}};}}

    \lstdefinestyle{searchstringstyle}{
	basicstyle=\ttfamily\scriptsize,
	captionpos=b,                    
	numbers=none,                    
	numbersep=4pt,                  
	showspaces=false,                
	showstringspaces=false,
	showtabs=false,                  
	tabsize=2,
	frame=single
}

\newtcolorbox{shadedbox}{
drop shadow southeast,
enhanced jigsaw,
colback=white,
boxrule=0.80pt,
left=0.3em,
right=0.3em,
top=0.1em,
bottom=0.05em
}

\lstdefinestyle{JavaStyle} {
  commentstyle=\color{mygreen},
  breakatwhitespace=false,
  keywordstyle=\color{violet},
  language=Java,
  frame=single,
  stringstyle=\color{blue},
  basicstyle=\scriptsize,
  captionpos=b,
  showstringspaces=false,
  tabsize=4, 
  numbers=left, 
  stepnumber=1, 
  numbersep=5pt, 
  numberstyle=\tiny\color{gray}, 
}

\lstdefinestyle{DSLStyle} {
  commentstyle=\color{mygreen},
  breakatwhitespace=false,
  keywordstyle=\color{violet},
  language=Java,
  frame=single,
  stringstyle=\color{blue},
  basicstyle=\scriptsize\ttfamily,
  captionpos=b,
  showstringspaces=false,
  literate={\ \ }{{\ }}1,
  tabsize=4, 
  numbers=left, 
  stepnumber=1, 
  numberstyle=\tiny\ttfamily,
  xleftmargin=2em,
  framexleftmargin=2em,
  morekeywords={GroupBias, highLevelDefinition, domain, education, source, WRONG_SAMPLING, sensitiveVariables, SensitiveVariable, name, gender, values, male, female, positiveOutcome, positive admission, unprivilegedGroup, SensitiveGroup, women, sensitiveValue, gender.female, privilegedGroup, men, gender.male, analysis, GroupAnalysis, scope, all people must have same admission probability despite gender, dataset, Dataset, id, admissions, predictedLabelName, admitted, filePath, admissions.csv, mappingOutcome, value, operator, EQUAL, datasetSentiveVariable, DatasetSensitiveVariable, sex, mappingSensitiveVariable, SensitiveVariableValue, mappingValue, datasetUnprivilegedGroup, datasetPrivilegedGroup, metric, Metric, StatisticalParity, toleranceValue, function, ExistingGroupFairnessMetric, optimalValue, arithmeticOperator, Operation, RATIO, leftSide, GroupSize, groupCondition, sensitiveGroup, AND, rigthSide, STATISTICAL_PARITY, relativeToDatasetSize, datasetSensitiveVariable, GREATER_EQUAL, MINOR_EQUAL, GREATER, mappingGroup, groundTruthLabelName,WRONG_ALGORITHM_BEHAVIOUR, rightSide,definition, HUMAN_DISCRIMINATION,Definition}
}

\lstdefinestyle{PythonStyle} {
  backgroundcolor=\color{white},   
  commentstyle=\color{deepgreen}, 
  breakatwhitespace=false,
  keywordstyle=\color{deepblue},
  language=Python,
  stringstyle=\color{deepgreen},
  basicstyle=\footnotesize\ttfamily,
  frame=single,
  captionpos=b,
  showstringspaces=false,
  numbers=left,
  numberstyle=\tiny\ttfamily,
  xleftmargin=2em,
  framexleftmargin=2em}

\definecolor{mygreen}{rgb}{0,0.6,0}
\definecolor{mygray}{rgb}{0.95,0.95,0.95}
\definecolor{myred}{rgb}{0.5,0,0}

\definecolor{verylightgray}{gray}{0.92}
\definecolor{ao(english)}{rgb}{0.0, 0.5, 0.0}

\newcommand{\SoBigDataITAck}{European Union - NextGenerationEU - National Recovery and Resilience Plan (Piano Nazionale di Ripresa e Resilienza, PNRR) - Project: “SoBigData.it - Strengthening the Italian RI for Social Mining and Big Data Analytics” - Prot. IR0000013 - Avviso n. 3264 del 28/12/2021\xspace}

\newcommand{\EmeliotAck}{EMELIOT national research project, which has been funded by the MUR under the PRIN 2020 program (Contract 2020W3A5FY)\xspace}

\newcommand{\FringeAck}{European Union – NextGenerationEU through the Italian Ministry of University and Research, Projects PRIN 2022 PNRR “FRINGE: context-aware FaiRness engineerING in complex software systEms" grant n.P2022553SL\xspace}
\usepackage[switch]{lineno}
\usepackage{orcidlink}

\begin{document}

\title{\revision{How Fair Are We? From Conceptualization To Automated Assessment of Fairness Definitions}}

\author{Giordano~d'Aloisio \orcidlink{0000-0001-7388-890X} \and
Claudio~Di~Sipio \orcidlink{0000-0001-9872-9542} \and
Antinisca~Di~Marco \orcidlink{0000-0001-7214-9945} \and
Davide~Di~Ruscio \orcidlink{0000-0002-5077-6793}
}

\authorrunning{d'Aloisio, Di Sipio, Di Marco, Di Ruscio} 

\institute{Giordano d'Aloisio \at
University of L'Aquila, Italy \\
\email{giordano.daloisio@univaq.it}
\and
Claudio {Di Sipio} \at
University of L'Aquila, Italy \\
\email{claudio.disipio@univaq.it}
\and
Antinisca {Di Marco} \at
University of L'Aquila, Italy \\
\email{antinisca.dimarco@univaq.it}
\and
Davide {Di Ruscio} \at
University of L'Aquila, Italy \\
\email{davide.diruscio@univaq.it}
}

\maketitle

\begin{abstract}
    
Fairness is a critical concept in ethics and social domains, but it is also a challenging property to engineer in software systems. With the increasing use of machine learning in software systems, researchers have been developing techniques to assess the fairness of software systems automatically. Nonetheless, many of these techniques rely upon pre-established fairness definitions, metrics, and criteria, which may fail to encompass the wide-ranging needs and preferences of users and stakeholders. To overcome this limitation, we propose a novel approach, called MODNESS, that enables users to customize and define their fairness concepts using a dedicated modeling environment. Our approach guides the user through the definition of new fairness concepts also in emerging domains, and the specification and composition of metrics for its evaluation \revision{through a dedicated Domain Specific Language}. Ultimately, MODNESS generates the source code to implement fair assessment based on these custom definitions.
In addition, we elucidate the process we followed to collect and analyze relevant literature on fairness assessment in software engineering (SE). We compare MODNESS with the selected approaches and evaluate how they support the distinguishing features identified by our study. Our findings reveal that \textit{i)} most of the current approaches do not support user-defined fairness concepts; \textit{ii)} our approach can cover additional application domains not addressed by currently available tools, e.g., mitigating bias in recommender systems for software engineering and Arduino software component recommendations; \textit{iii)} MODNESS demonstrates the capability to overcome the limitations of the only two other Model-Driven Engineering-based approaches for fairness assessment.
\end{abstract}


\section{Introduction}
\label{sec:introduction}
In recent years, \textit{machine learning} (ML) has become increasingly widespread and popular. ML is the backbone of many systems we use daily, such as intelligent speakers and e-commerce systems that provide personalized recommendations. However, creating data-driven decision-making systems using ML can be complex due to the needed vast amount of training data, the complexity of the information relationships, and continuous learning activities that aim to improve the accuracy of such systems \cite{giray_software_2021,nahar_meta-summary_2023}. Unfortunately, these factors make it challenging for developers to ensure that these systems are free from \textit{bias}. To address this issue, the scientific community has begun evaluating ML-intensive systems in terms of quality attributes such as \textit{explainability}, \textit{privacy}, and \textit{fairness} \cite{giray_software_2021,villamizar_requirements_2021,muccini_software_2021}. Among these quality attributes, fairness gained considerable relevance in the SE community \cite{giray_software_2021,villamizar_requirements_2021,muccini_software_2021,kumeno_sofware_2020,zhang_machine_2020}. 
The relevance of fairness is also highlighted by the \textit{AI Act} \revision{recently approved by the European Commission, where it is described as a key quality property for \textit{high-risk} AI-enabled systems.\footnote{\url{https://digital-strategy.ec.europa.eu/en/policies/regulatory-framework-ai}}}

Fairness (and relative unbias) can be defined as \textit{"the absence of any prejudice or favouritism toward an individual or group based on their inherent or acquired characteristics"} \cite{mehrabi_survey_2021}. Although the concept of \textit{fairness} has been introduced primarily in the context of machine learning (ML) systems (mainly for legal reasons \cite{mehrabi_survey_2021,caton_fairness_2020}), the concept of \textit{bias} originates in the ethical domain (referring to the general concept of \textit{discrimination}), and has been adapted to the AI and ML domain \cite{olteanu2019social}.  The relevance of software fairness in ML-intensive systems has also been made popular by infamous incidents in the recruitment instrument employed by Amazon \rev{which penalized women candidates for IT job positions},\footnote{\revision{\url{https://www.bbc.co.uk/news/technology-45809919}}} and the criminal recidivism predictions made by the commercial risk assessment software COMPAS \rev{which misjudged black individuals based on biased profiling} \cite{angwin_machine_2016}. In general, an ML system is said to be \textit{biased} (or \textit{unfair}) if the output of the ML model is directly correlated to the value of a set of so-called \textit{sensitive variables}, like the race or gender of a person. Starting from a definition of bias for a particular domain, several fairness analyses can be computed by selecting a set of metrics compliant with a particular scope. The process of analyzing if a given system produces fair outcomes involves four steps \cite{mehrabi_survey_2021,manila2023}: \emph{bias definition}, \emph{fairness analysis specification}, \emph{analysis implementation}, and \emph{fairness assessment}. However, performing these steps manually can be challenging and error-prone, \revision{requiring actors with different levels and domains of expertise}. \revision{Moreover, several works have highlighted the need of \emph{democratising} the development and quality assessment (including fairness) of ML-based systems through automated frameworks and improving the link between domain and technical experts \cite{sundberg_democratizing_2023,manila2023,10.1145/3550355.3552401,10.1145/3411764.3445261}.} 

To address this issue, several toolkits and frameworks have been developed to automate the fairness analysis of ML-based systems by \textit{i)} identifying the sources of bias and \textit{ii)} suggesting appropriate countermeasures. These tools are helpful, but they have some limitations: they rely on predefined fairness definitions that may not suit every application domain, and they do not allow users to customize the concepts of bias, fairness, and the corresponding metrics. \rev{The limitations of existing approaches are highlighted in this paper through a lightweight literature survey and discussion.}

As recent studies and surveys have shown \cite{mehrabi_survey_2021,caton_fairness_2020}, there is no consensus on defining and measuring fairness across different domains. Therefore, there is a need for dedicated languages that enable users to specify their fairness criteria and metrics \cite{10.1145/3411764.3445261}. Furthermore, while different types of bias have been studied in specific domains (such as recommender systems \cite{deldjoo2022fairness,wang_survey_2023}), ethical, legal, or social factors still influence the notion of bias.
 
In this paper, we introduce \tool, a versatile tool that helps to conceptualize, personalize, and assess fairness definitions across different domains. It is based on the Model-Driven Engineering (MDE) paradigm and features a specialized metamodel that caters to two levels of abstraction: the \textit{conceptual} and the \textit{development} level. With the former, users can define critical fairness concepts such as bias and metrics to detect unfairness independently of the application domain. The latter utilizes such a definition to evaluate and generate code modules using real-world datasets automatically. \revision{In addition, we provide a tailored Domain Specific Language (DSL) for the specification of models compliant with the proposed metamodel.}

\subsection{\revision{Research Questions}}

\revision{In this paper, we answer the following three research questions (RQ):}

\revision{
\begin{itemize}
    \item \rqfirst 
    \item \rqsecond 
    \item \rqthree
\end{itemize}
}

\revision{To address RQ$_1$, we conduct a lightweight literature review by carefully inspecting existing tools and approaches collected from top SE venues according to specifically defined inclusion and exclusion criteria.
Then, the elicited approaches have been characterized in terms of recurrent operations in the fairness assessment workflow, \ie bias definition, fairness analysis specification, and implementation and assessment.} 

\revision{To tackle RQ$_2$, we evaluate the \textit{expressiveness} and \textit{accuracy} of \tool. We gauge \textit{expressiveness} by demonstrating \tool's capability to model and implement a diverse range of use cases concerning bias and fairness analysis. Conversely, we assess \textit{accuracy} by verifying that the code generated by \tool can reliably identify biases in the chosen use cases. In particular, we use \tool to model and generate the implementation of the most notable use cases in the fairness domain (\ie education \cite{austin_will_2016}, social \cite{angwin_machine_2016,kohavi_scaling_nodate} and financial \cite{hofmann2013uci,moro2014data}). In addition, we show how \tool can cover two additional non-traditional use cases in the fairness domain, adopting metrics not common in the fairness literature. The former is about a recent study to evaluate popularity bias in recommender systems for software engineering (\textit{TPL}) \cite{nguyen2023dealing}. The latter covers recommendations for Arduino software components \cite{di2023resyduo}. Moreover, we employ two use cases, \ie a traditional use-case about bias in university admissions (\textit{University}) and the \textit{TPL} use-case presented above as running examples to explain the key \tool concepts throughout the paper.} 


 
\revision{To address RQ$_3$, we compare \tool with two baselines elicited from the literature review conducted to answer the RQ$_1$ \ie FairML \cite{10.1145/3550355.3552401} and MANILA \cite{manila2023}. The rationale behind this choice is that they \textit{i)}  have been developed using the MDE paradigm and \textit{ii)} allow the specification of the fairness assessment workflow. Assessing the quality of MDE-based tools is daunting since they usually rely on tailored metamodels conceived for a specific application domain. Prior works have defined a set of quality metrics that investigate several aspects such as expressiveness, completeness, or portability\cite{Bertoa2010QualityAF}. Within the scope of our paper, we establish two dimensions for facilitating comparison: \textit{expressiveness} and \textit{automation}.
}

\subsection{\revision{Paper Contributions}}

The main contributions of the paper are the following:

\begin{itemize}
    \item We carefully review existing tools and approaches that automatically detect group and individual bias;
    \item We propose a tailored metamodel \revision{and a corresponding DSL}\footnote{The full source code, as well as the replication package of the performed experiments, are available at \cite{Daloisio_MODNESS_2024}} that is capable of representing existing fairness definitions and providing an extensible mechanism to conceptualize new ones;
    \item We evaluate the proposed approach by modeling a set of use cases belonging to different application domains, \ie social, financial, software development, and IoT;
    \item We compare the proposed approach with two MDE-based baselines in terms of quality aspects considered, highlighting how \tool can overcome their current limitations .
\end{itemize}
This paper is organized as follows: Section \ref{sec:motivation} provides an overview of terminology and the particular phases of fairness assessment workflows. Section \ref{sec:slr} reviews existing approaches and tools for fairness analysis presented in top SE venues. The \tool approach and supporting tools are presented in Section \ref{sec:proposedApproach}. 
In Section \ref{sec:discussion}, we show applications of \tool on different use cases. Threats to the validity of the performed evaluation are discussed in Section \ref{sec:threats}. Section \ref{sec:conclusion} concludes the paper and describes future directions.

\section{\revision{Fairness Assessment: Key Concepts and a General Workflow}}
\label{sec:motivation}
In this section, we recall the main concepts of bias and fairness and describe a general process for fairness assessment. \revision{Additionally, we introduce two case studies that we use throughout the paper to illustrate the fundamental concepts and underlying workflow of \tool. These examples are deliberately chosen from significantly different domains to highlight the heterogeneity of key fairness concepts:} 

\begin{enumerate}
    \item \noindent\revision{\textbf{University Admission (\textit{University}).} It was originally presented in \cite{austin_will_2016} and involves a university that uses an ML-based system to determine student admissions. It is necessary to ensure that the system is fair. Specifically, the concern is whether or not the system is biased against \textit{women}.}

    \item \noindent\revision{\textbf{Popularity Bias in TPL Recommendation (\textit{TPL}).} It concerns the problem of \textit{popularity bias} in Third-Party Library (TPL) recommendations \cite{nguyen2023dealing}. TPL recommendation concerns the development of recommender systems that can suggest TPL to developers based on the software they are developing. In the original paper, authors highlight how very popular libraries (\ie with high \textit{``reputation"}) are more likely to be recommended mainly because several developers are using them. On the contrary, more specific or recent libraries are less likely to be recommended, even if they are more appropriate to the development task at hand.}
\end{enumerate}



\begin{figure}[ht!]
    \centering
    \includegraphics[width=\linewidth]{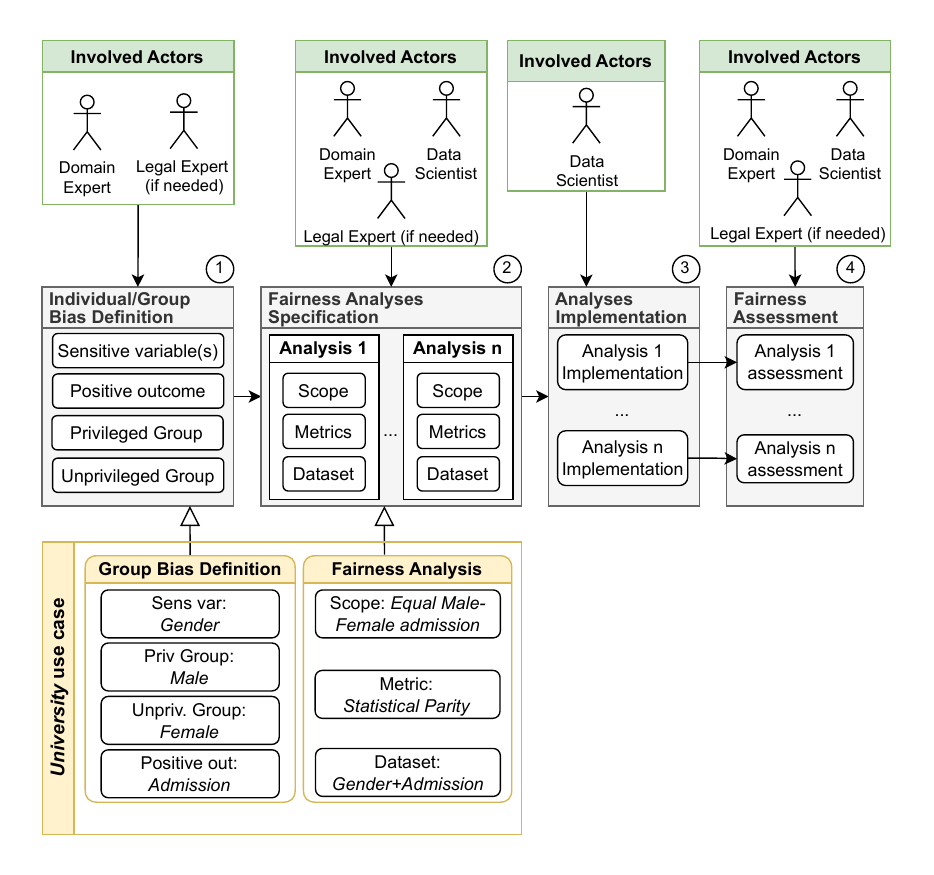}
    \caption{\rev{General fairness assessment workflow. On top, there are the main actors involved in each step, while on the bottom, there is the instantiation of the key concepts for the \textit{Univerisity} use case.}}
    \label{fig:workflow}
\end{figure}

\revision{Figure \ref{fig:workflow} illustrates a general fairness assessment workflow, highlighting the main concepts at each step. Additionally, the top of the figure shows the main actors involved in each step, while the bottom provides an instantiation of the key concepts for the \textit{University} use case.} The workflow has been derived by analysing the behaviour of some of the most adopted fairness toolkits and libraries (\ie Aequitas \cite{10.1145/3238147.3238165}, IBM AIF360 \cite{bellamy_ai_2019}, and Fairlearn \cite{bird2020fairlearn}) and by reviewing foundational papers on bias and fairness \cite{mehrabi_survey_2021,brun_software_2018,chen_fairness_2022}. 
When defining bias and conducting fairness analysis, it is important to consider the specific domain being studied, as highlighted in recent research \cite{mehrabi_survey_2021,caton_fairness_2020}. 

The fairness assessment process may involve multiple parties, with three key actors typically identified:
\begin{itemize}
    \item[--] \textbf{Domain expert} provides domain-specific knowledge, \revision{namely the person in charge of managing admissions to the university or the \rev{developer of the recommender system in our examples}};
    \item[--] \textbf{Data scientist} provides knowledge about fairness metrics and their implementation\revision{, namely  the data scientist provides knowledge about methods and metrics to assess possible bias in the university admission and recommender systems, based on the bias definitions provided by the domain expert in our examples};
    \item[--] \textbf{Legal expert} provides knowledge about specific regulations if needed in the fairness assessment process\revision{, namely the legal expert provides knowledge about possible regulations concerning university admissions or recommender systems for particular domains in our examples}. 
\end{itemize}  

The first step in the workflow depicted in Fig. \ref{fig:workflow} is the \textit{bias definition}, which involves both the domain and legal experts \revision{(step \circledw{1} in the Figure)}. 
In general, bias can be of two types: \cite{mehrabi_survey_2021,caton_fairness_2020}: 
\begin{itemize}
    \item[--] \textbf{Group bias:} if discrimination \rev{or favoritism} is assessed 
    at the group level \revision{(\eg all the \textit{women} for the \textit{University} use case, or all the \textit{less popular} libraries for the \textit{TPL} use case)}.
    \item[--] \textbf{Individual bias:} if discrimination \rev{or favoritism} is assessed for any individual\rev{, regardless of their belonging to a specific group}. For instance, in our use cases, \rev{a student must not be admitted to the university only because they are a relative of the university rector, or a library must not be recommended only based on previous user preferences (regardless of its relevance).}
\end{itemize}
\revision{It is worth mentioning that, even if considered by the literature, use cases regarding \textit{individual bias} are less common and, in general, \textit{individual bias} is more difficult to mitigate compared with \textit{group bias} \cite{mehrabi_survey_2021,chen_fairness_2022,hort_bias_2023}.}

To establish a definition of bias within a specific domain, it is crucial to pinpoint the following key concepts, as outlined in \cite{mehrabi_survey_2021}: 
\begin{itemize} 
    \item[--] \textbf{Sensitive variables:} these are the variables that have the potential to lead to discrimination;
    \item[--] \textbf{Positive outcome:} This refers to a specific prediction generated by the ML system that could potentially result in discriminatory consequences. \revision{Note that this concept is domain-dependent and may change based on the perspective from which we observe a given use case (e.g., a prediction about a customer positively subscribing to a bank term deposit may be good for the business but not for the customer's wallet)} 
    \item[--] \textbf{Privileged} and \textbf{Unprivileged} groups: These are the groups or entities identified based on specific values of the sensitive variables. Privileged groups \revision{may be} favored by the system, while unprivileged groups \revision{may} face discrimination. \revision{Note how these two groups are often referred to as \textit{sensitive groups}, i.e., groups that must be protected from discrimination or favoritism (e.g., by some regulations).}
\end{itemize}

\revision{Concerning the \textit{University} use case (as reported at the bottom of Fig. \ref{fig:workflow})}, domain and legal experts want to assess if the ML system under analysis discriminates against women. Thus, the type of examined bias is \textit{group bias}, with the sensitive variable being \textit{gender}. A positive outcome would be \textit{\revision{successful} admission} to the university, and the privileged group is \textit{men} while the unprivileged one is \textit{women}.
\revision{For the \textit{TPL} use case, domain experts want to ensure that not only popular libraries are recommended. Hence, the type of bias is still \textit{group bias}, where the sensitive variable is \textit{popularity}. A positive outcome is a \textit{recommendation} from the system. The privileged group is \textit{popular libraries} while the unprivileged group is \textit{unpopular libraries}.}

From an abstract bias definition, several fairness analyses can be depicted \revision{(step \circledw{2} in Fig \ref{fig:workflow})}. In general, a fairness analysis has a \textit{scope} (\ie a fairness definition), a set of \textit{metrics} compliant to the given scope (which can be fairness metrics known in the literature or can be custom ones), and has a \textit{dataset}, which contains all the information needed to perform the given analysis.

\revision{In the considered \textit{University} scenario (see Fig. \ref{fig:workflow}), the domain and legal experts decide that the ML system is fair if \textit{men} and \textit{women} have the same probability of being admitted to the university. This is the scope of the analysis.} Hence, the data scientist suggests using the \textit{Statistical Parity} fairness metric \cite{kusner_counterfactual_2017}, a metric compliant with the given fairness definition. Finally, the domain expert and the data scientist collect all the needed information (\ie the predictions of the ML model and the gender of each person) inside a dataset that will be used for the analysis. 
It is important to mention that other analyses can be conducted. For example, domain and legal experts may want to examine whether someone is only admitted to a university based on their high school grades, regardless of gender. Hence, other metrics (\eg \textit{Equalized Odds} \cite{hardt_equality_2016}) can be used to cover this fairness definition, and further information has to be provided in the dataset.

\revision{\rev{For the \textit{TPL} use case, domain experts can state that a recommender system is free from popularity bias if relevant libraries are recommended despite their low popularity \cite{nguyen2023dealing}}. To assess this definition, they adopt a variation of the \textit{Coverage} metric, which measures the percentage of \textit{non-popular} items recommended over the total recommendations \cite{NGUYEN2019110460}. Note how this metric does not come from the fairness literature but is an adaptation of a metric from the recommender systems domain to assess popularity bias \cite{nguyen2023dealing}. Finally, like for the \textit{University} use case, a dataset containing the recommended libraries and their popularity is collected to perform the fairness assessment.}

The next step in the fairness assessment process is the implementation of the defined fairness analyses \revision{(step \circledw{3} in Fig. \ref{fig:workflow})}. Concretely, this means implementing an automatic procedure that, given a specific dataset as input, computes all the selected metrics and returns the calculated values. \revision{In our examples}, the data scientist has to implement software using a programming language (\,  e.g. Python) that takes as input the dataset and computes the \textit{Statistical Parity} \revision{or \textit{Coverage} fairness metrics}.

Once the results are computed, all parties must evaluate them to determine the system's fairness \revision{(step \circledw{4} in Fig. \ref{fig:workflow})}. For example, \revision{in our scenarios}, the data scientist notes that a \textit{Statistical Parity} score of zero \revision{or a \textit{Coverage} score of one} indicates fairness. Next, legal and domain experts analyse the results to assess whether the ML system is fair based on this definition.

The process depicted in Fig. \ref{fig:workflow} and described earlier can be challenging and prone to mistakes, mainly because it involves several stakeholders and interconnected tasks. To address these issues, we introduce a tool-based approach that automates the assessment of fairness. This approach enables users to define their own notions of bias and fairness, enlarging the applicability of fairness assessment to multiple application domains. We present the details of this approach in Section \ref{sec:proposedApproach}.

\section{Existing Approaches for Fairness Assessment}
\label{sec:slr}
This section presents the procedure exploited to gather pertinent literature within the realm of fairness assessment. First, we describe the adopted procedure in Section \ref{sec:methodology}, including the search string and the inclusion and exclusion criteria. Subsequently, Section \ref{sec:feature} delves into an exploration of key features and sub-features constituting the fairness assessment workflow that we use to classify the selected approaches, while Section \ref{sec:tools} provides an overview and classification of the collected approaches.

\subsection{Methodology} \label{sec:methodology}


In this section, we provide an overview of prominent approaches within the domain of bias and fairness assessment in ML-based systems focusing on the SE community. Please note that our aim is not to present a comprehensive survey of the entire field, as it goes beyond the scope of this paper. Instead, we have adopted a \revision{tool-supported procedure} inspired by the well-established "four W-question strategy" \cite{10.1016/j.infsof.2010.12.010} to select existing approaches that perform bias detection 
using automated or \revision{tool-supported} methods. 
 \revision{In particular, our analysis is confined to peer-reviewed scientific works that make significant contributions in two key areas: \textit{i)} defining or addressing fairness concerns within software systems, and \textit{ii)} employing automated methods to mitigate identified biases while considering notable datasets.}

The four W-questions guiding our approach are as follows:

\begin{itemize}
  \item \emph{Which?} We conducted a comprehensive search, combining both automated and manual methods, to gather relevant papers from a variety of sources, including conferences and journals. 
  \item \emph{Where?} \revision{Our literature analysis focused on prominent software engineering venues, encompassing eleven conferences: ASE, ESEC/FSE, ESEM, ICSE, ICSME, ICST, ISSTA, MSR, SANER, FASE, and MODELS as well as five journals: EMSE, IST, JSS, TOSEM, and TSE. In particular, we collect relevant information for those venues, \ie title and abstract, that we used in the filtering process.} 
  To automate this process, we utilized the \textit{Scopus} database\footnote{\url{https://scopus.com}} and employed advanced search and export functions to retrieve all papers published in specific venues within the temporal range we decided. 
  \item \emph{What?} For each article, we extracted information from the title and abstract by applying predefined keywords to ensure relevance to our research focus.
  \item \emph{When?} Given that automated fairness assessment is a relatively recent research area, our search was limited to the most recent five years, spanning from 2017 to 2023. This temporal constraint allowed us to capture the latest developments and trends in the field. It is worth noticing how the query was executed in April of 2024, hence 2024 has not been considered.
\end{itemize}

%

\begin{table}[h!]
	\centering
	\footnotesize
 \revision{
	\caption{\revision{Number of papers for the related topics.}}\label{tab:slr-results}
	\begin{tabular}{|l|c|c|c|}
		\hline
		& FAIR & ML & TOOL \\ \hline
		FAIR & 241 & \cellcolor{gray!25} & \cellcolor{gray!25} \\ \hline
		ML & \cellcolor{green!25}51 & 1,931 & \cellcolor{gray!25} \\ \hline
		TOOL & \cellcolor{green!25}56 & 123 & 3,438 \\ \hline
	\end{tabular}}
\end{table}


\revision{We export the relevant papers from Scopus and exploit dedicated Python scripts to search in title and abstract the following set of keywords in \textit{AND} conjunction:}

\emph{(i)} \textbf{FAIR}: ``\emph{fairness}'' or ``\emph{bias}''; 
\emph{(ii)} \textbf{ML}: ``\emph{data science}'' or ``\emph{machine learning}; \emph{(iii)} \textbf{TOOL}: ``\emph{toolkit},'' or ``\emph{definition}'', or ``\emph{audit}'', or ``\emph{testing}'' or ``\emph{model-based}''.
Table~\ref{tab:slr-results} reports the number of papers that contain such keywords in the corresponding column and row (e.g., 123 papers contain at least one term belonging to ML and TOOL sets). Our ideal targets are papers containing at least one keyword for all the defined sets of terms, \ie FAIR, ML, and TOOL. By running such a combination, we obtain only 15 works considering the abovementioned criteria. \revision{Therefore, we enlarged the set of eligible papers to two additional combinations highlighted in green, \emph{(i)} FAIR and ML; or \emph{(ii)}  FAIR and TOOL. In the end, we obtained a total of 107 papers, including duplicate papers. By removing those ones, we ended up with 61 scientific papers, including journal and conference publications.}

Starting from this initial set of works, we manually inspected the title and abstract to scale down the search to meet our requirements. In particular, we defined the following inclusion and exclusion criteria:

\smallskip
\noindent
\ding{52} \textbf{Inclusion criteria:} We included all the approaches that use traditional bias definitions, \ie group or individual. Furthermore, we consider toolkits or frameworks that provide an automatic or semi-automatic strategy to assess fairness on a set of use cases. Some of them have been published as extensions of initial works. Therefore, we consider the most recent version of the tool in such cases. 

\smallskip
\noindent
\ding{54} \textbf{Exclusion criteria:} The study we conducted intentionally excludes foundational papers that primarily provide a high-level abstract definition of fairness, such as surveys \cite{mehrabi_survey_2021,caton_fairness_2020}, empirical studies \cite{10.1145/3597503.3639083}, or position papers \cite{brun_software_2018}. \revision{Additionally, we excluded papers focusing on improving the fairness of underlying ML models \cite{daloisio_debiaser_2023,kamiran_data_2012}, as our focus is on automating the assessment process through the explicit specification of the application domain.}

\begin{figure}
    \centering
    \includegraphics[width=\linewidth]{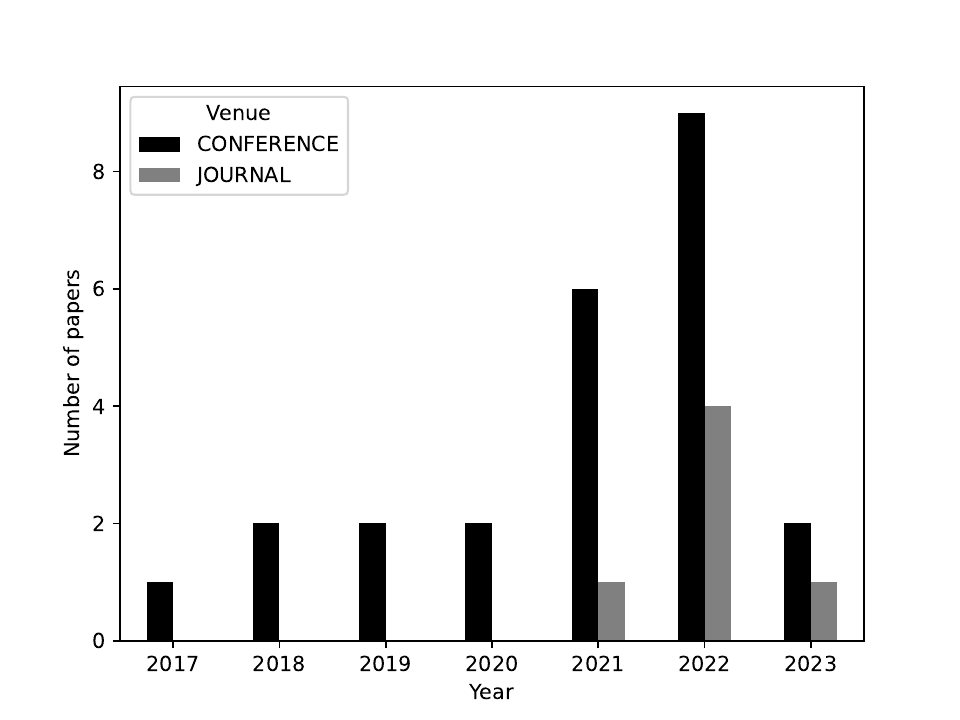}
    \caption{\revision{Number of selected papers per year.}}
    \label{fig:plot}
\end{figure}

To ensure an unbiased selection process, we employed a rigorous approach. Two different authors independently evaluated all the papers, and the two senior co-authors thoroughly reviewed the entire selection process. \revision{ Ultimately, this meticulous process yielded a total of 26 works.\footnote{\rev{The complete list of selected and excluded papers with the venues name is available in the replication package \cite{Daloisio_MODNESS_2024}}} Figure \ref{fig:plot} depicts the retrieved papers divided by year. Notably, there is an increasing trend with a peak in 2022, indicating that automating fairness assessment is becoming increasingly relevant in the SE community. The rising number of journal publications confirms this, suggesting that researchers share more mature results than the initial studies that appeared in 2017.} 
\revision{However, there is a noticeable decrease in the number of published papers in 2023, although this trend represents frameworks selected using the abovementioned procedure. Therefore, it is not representative of the whole trend in SE concerning fairness assessment.}



\subsection{Elicited features} \label{sec:feature}

Table \ref{tab:toolComparison} summarises the list of papers we collected by means of the previously described process. For each approach, we list the name of the tool, the venue, the year, and the underpinning mechanism used to define and assess fairness (if any). Furthermore, starting from the four steps of the general fairness assessment workflow described in Section \ref{sec:motivation}, we elicit six different features to evaluate the degree of automation and customization of the selected approaches. These features are grouped in Table \ref{tab:toolComparison} by the primary step of the workflow described in Figure \ref{fig:workflow} they belong to, \ie \textit{bias definition}, \textit{fairness analysis specification}, \textit{analysis implementation \& fairness assessment}. The selected features are as follows:

\noindent \ding{228}  \textbf{F1 - Bias definition}: The approach models and assesses individual bias definitions, group bias definitions, or both;

\noindent \ding{228} \textbf{F2 - \rev{Domain} bias definition}: The approach implements an extension mechanism to provide a bias definition \revision{that is tailored for a specific domain and agnostic from a specific fairness analysis or dataset (see steps \circledw{1} and \circledw{2} of Fig \ref{fig:workflow})};

\noindent \ding{228} \textbf{F3 - Custom metric definition}: Similar to the previous one, the approach allows the definition of additional metrics to detect bias \revision{(\ie metrics for less common use cases like \textit{TPL})};

\noindent \ding{228}  \textbf{F4 - Metric composition}: It is possible to combine defined metrics to create new ones, \revision{for instance, by means of aggregation functions};

\noindent \ding{228} \textbf{F5 - Automated fairness assessment}: The underlying system assesses the fairness by automatically generating the corresponding source code;  

\noindent \ding{228} \textbf{F6 - Tool availability}: The paper is supported by a publicity available tool;

\revision{For each tool, we marked these features with \textit{supported} (\tick)  in Table \ref{tab:toolComparison} while we left blank unsupported features.}
Concerning the feature \textit{fairness definition}, the symbols \circled{I} and \circledb{G} are used for \emph{individual} and \emph{group} bias, respectively.


\begin{table*}[htb!]
    \centering
    \caption{Comparison of the existing fairness toolkit and approaches.}
    \label{tab:toolComparison}
    \resizebox{\textwidth}{!}{
    \begin{tabular}{|l|l|c|c| c |c |c | c| c|}
    \hline
        
        \rowcolor{lightgray} 
        \multicolumn{3}{|c|}{} & \multicolumn{2}{|c|}{\textbf{Bias Definition}} & \multicolumn{2}{|c|}{\Gape[0pt][2pt]{\makecell[c]{\textbf{Fairness Analysis}\\\textbf{Specification}}}} & \multicolumn{2}{|c|}{\Gape[0pt][2pt]{\makecell[c]{\textbf{Analysis Implementation \&}\\\textbf{Fairness Assessment}}}} \\ \hline
        
        \textbf{Approach} & \textbf{Venue \& Year} & \textbf{Base strategy} & \textbf{F1 - Bias def.} & \makecell[c]{\textbf{F2 - \revision{Abstract}}\\\textbf{bias def.}} & \makecell[c]{\textbf{F3 - Custom}\\\textbf{metric def.}} & \makecell[c]{\textbf{F4 - Metric}\\\textbf{Comp.}} & \makecell[c]{\textbf{F5 - Automated}\\ \textbf{fairness assess.}} & \textbf{F6 - Tool avail.} \\ \hline

       Aequitas \cite{10.1145/3238147.3238165} & ASE (2018) & Search-based & \circledb{G} &  &  &  & \tick & \tick \\ \hline
       
        Themis \cite{angell_themis_2018} & ESEC/FSE (2018) & Search-based & \circledb{G} &   & \tick &   & \tick & \tick \\ \hline
       
        TILE \cite{sharma_testing_2019} & ICST (2019) & Metamorphic testing & \circled{I} &   &   &   & \tick &   \\ \hline
        
        ADF \cite{10.1145/3377811.3380331} & ICSE (2020) & Adversarial DL & \circled{I} &   &   &   &   & \tick \\ \hline
        
        Fairway \cite{chakraborty_fairway_2020} & ESEC/FSE (2020) & Search-based & \circledb{G} &   & \tick &   & \tick & \tick \\ \hline
        
        DeepInspect \cite{10.1145/3377811.3380400} & ICSE (2020) & Deep learning & \circledb{G} &   &   &   & \tick & \tick \\ \hline
        
        AITEST \cite{aggarwal_testing_2021} & ICSE (2021) & Search-based & \circled{I} &   &   &   & \tick &   \\ \hline
        
        EIDG \cite{zhang2021efficient} & ISSTA (2021) & Search-based & \circled{I} &   &   &   &   & \tick \\ \hline
        
        Fair-SMOTE \cite{10.1145/3468264.3468537} & ESEC/FSE (2021) & Situation testing & \circled{I},\circledb{G} &   &   &   & \tick & \tick \\ \hline

     \revision{Biswas and Rajan \cite{biswas_fair_2021}} & \revision{ESEC/FSE (2021)} & \revision{Casual fairness}  & \textcolor{blue}{\circled{G}} & \revision{} & \revision{} & \textcolor{blue}{} & \revision{\tick} & \revision{\tick} \\    \hline
        
        Fairea \cite{10.1145/3468264.3468565} & ESEC/FSE (2021) & Mutation testing & \circledb{G} &   &   &   & \tick & \tick \\ \hline
        
        BiasFinder \cite{9653830} & TSE (2021) & Mutation testing & \circled{I} &   &   &   & \tick & \tick \\ \hline
        
        FairKit-learn \cite{9793775} & ICSE (2022) & Search-based & \circled{I},\circledb{G} &   &   &   & \tick & \tick \\ \hline
        
        PAIRFAIT-ML\cite{10.1145/3510003.3510202} & ICSE (2022) & Search-based & \circledb{G} &   &   &   &   & \tick \\ \hline

        MAAT\cite{chen_maat_2022} & ESEC/FSE (2022) & Ensemble learning & \circledb{G} &   &   &   & \tick & \tick \\ \hline

        FairMask\cite{peng_fairmask_2023} & TSE (2022) & Hybrid  & \circledb{G} &   &   &   & \tick & \tick \\ \hline
        
        ExpGA \cite{10.1145/3510003.3510137} & ICSE (2022) & Genetic algorithm & \circled{I} &   &   &   & \tick & \tick \\ \hline
        
        Astraea \cite{soremekun_astraea_2022} & TSE (2022) & Grammar-based gen. & \circled{I},\circledb{G} & \tick &   &   & \tick &   \\ \hline
        
        SBFT \cite{perera_search-based_2022} & EMSE (2022) & Genetic algorithm & \circled{I} &   &   &   &   & \tick \\ \hline
        
        NeuronFair \cite{9793943} & ICSE (2022) & Adversarial DL & \circled{I} &   &   &   & \tick & \tick \\ \hline
        
        LTDD \cite{10.1145/3510003.3510091} & ICSE (2022) & Linear regression & \circledb{G} &   &   &   & \tick & \tick \\ \hline
        
        iRec2.0 \cite{wang_context-_2022} & TOSEM (2022) & Optimization problem & \circled{I} &   &   &   &   & \tick \\ \hline

        FairML \cite{10.1145/3550355.3552401} & MODELS (2022) & MDE-based & \circled{I},\circledb{G} & \tick  &   &   & \tick & \tick \\ \hline

\revision{AequeVox} \cite{rajan_aequevox_2022} & \revision{FASE (2022)} & \revision{Metamorphic testing} & \textcolor{blue}{\circledb{G}} & \textcolor{blue}{} & \textcolor{blue}{} & \textcolor{blue}{} & \revision{\tick} & \revision{\tick} \\ \hline

        MANILA \cite{manila2023} & FASE (2023) & MDE-based & \circled{I},\circledb{G} &   &   & \tick & \tick & \tick \\ \hline




\revision{FairiFy \cite{biswas_fairify_2023}} & 
\revision{ICSE (2023)} & \revision{Satisfiability modulo theories} &  \revision{\circled{I}} &   
 &   &   \revision{} & \revision{\tick} & \revision{\tick} \\ \hline

\revision{DICE \cite{monjezi_information-theoretic_2023}} & 
\revision{ICSE (2023)} & \revision{Search-based testing} &  \revision{\circled{I}} &  &  &  &   
\revision{\tick} & \revision{\tick} \\ \hline

    \end{tabular}
   }
\end{table*}

\subsection{Selected approaches} \label{sec:tools}


Aequitas \cite{10.1145/3238147.3238165} exploits three different search-based strategies to assess the group fairness of benchmarking ML-based classifiers, \ie random, semi-directed, and fully directed. The results of the conducted evaluation show that Aequitas can reduce the unfairness of the examined ML models. 

Similarly, Themis \cite{angell_themis_2018} provides a GUI to specify the schema of the dataset on which a user wants to assess fairness and generates a test case for it. Furthermore, it generates a report showing the relative group fairness for each variation in the value of the variables.


Sharma \etal \cite{sharma_testing_2019} investigate fairness in the learning phase of an ML algorithm. The proposed tool, called TILE, relies on a metamorphic testing approach to analyze the so-called \textit{balanced data usage}, \ie the learner should treat all data in the training set equally. TILE assesses fairness regarding this metric by being tested on several \texttt{scikit-learn} ML models. 


A scalable gradient-based algorithm called Adversarial Discrimination Finder (ADF) has been proposed to assess individual bias by injecting individual discriminatory instances into a given dataset \cite{10.1145/3377811.3380331}. The global generation phase generates discriminatory entities by combining generative models and clustering techniques. Such data are refined by the local generation phase using underpinning gradients. As stated in the evaluation, the ADF algorithm overcomes two state-of-the-art tools regarding effectiveness and efficiency.

Fairway is a tool proposed by Chakraborty \etal that covers both bias detection and mitigation \cite{chakraborty_fairway_2020}. This tool works under the \textit{Equal Opportunity} (EO) \cite{hardt_equality_2016} and \textit{Average Odds} (AO) \cite{berk_fairness_2018} group definitions of fairness by identifying \textit{ambiguous} data points. Next, it removes the bias learned by the ML algorithm through an optimization approach. Fairway succeeds in improving fairness under the EO and AO definitions.


Fairness in image classification has been investigated in \cite{10.1145/3377811.3380400}. The authors propose DeepInspect, a deep learning approach to mitigate two types of discrimination, \ie confusion and bias. The underpinning network uses the neuron activation probability (NAP) matrix to predict the abovementioned discriminations. The results show that DeepInspect performs better than existing approaches in terms of accuracy, thus detecting misclassification correctly. 

AITEST \cite{aggarwal_testing_2021} is a tool that combines constraint-based linear optimization with the local interpretable model-agnostic explanation (LIME) techniques to perform individual fairness assessment. The proposed hybrid search strategy outperforms two notable fairness toolkits, \ie Themis and Aequitas.

The same authors of \cite{10.1145/3377811.3380331} extend their former work by proposing an \textit{Efficient Individual Discriminatory Instances Generator} (EIDIG) to generate individual fairness test cases for DNN models systematically \cite{zhang2021efficient}. The evaluation demonstrates that considering the gradient of the model output instead of the gradient of the loss improves the ADF's overall accuracy and F1 scores. 

Fair-SMOTE \cite{10.1145/3468264.3468537} has been conceived to remove bias by exploiting a relabeling strategy. After the bias detection phase, it rebalances sensitive groups using the K-nearest neighbour algorithm. The results show that Fair-SMOTE solved biases before training the models compared to existing approaches.

\revision{Biswas and Rajan \cite{biswas_fair_2021} apply fairness assessment to the preprocessing steps of ML pipelines. Built on top of the fairness causality definition, the approach automatically computes the fairness metrics at each preprocessing step of a given ML pipeline.} 

Fairea \cite{10.1145/3468264.3468565} mitigates group biases based on bias-mitigation models generated using a mutation engine. The tool identifies and tests five bias mitigation strategies to measure the trade-off between accuracy and fairness. Fairea has been evaluated using two different metrics, \ie statistical parity difference and average odds difference.

Similarly, BiasFinder \cite{9653830} adopts mutation testing to assess fairness in sentiment analysis (SA) systems. The mutant engine produces actual instances by relying on bias-targeting templates extracted from the textual content. The tool was evaluated quantitatively and qualitatively by considering two large SA datasets and human annotators. 

Being built on top of sklearn and AIF360 frameworks, the Fairkit-learn toolkit \cite{9793775} provides a comprehensive platform to train, test, and compare ML models by considering fairness aspects. The tool retrieves fairer models than the two abovementioned libraries by relying on a set of Pareto-optimal strategies. 

PAIRFAIT-ML \cite{10.1145/3510003.3510202} exploits three different dynamic search algorithms to support hyper-parameters tuning by providing a set of bias-free configurations. The evaluation shows that the retrieved items reduce the bias by considering two metrics, \ie equal opportunity and average odd difference.

Chen \etal \cite{chen_maat_2022} propose MAAT, an ensemble approach to optimize the bias removal by combining two different models, \ie fairness and performance models. The former relies on the undersampling strategy to mitigate the group bias. The latter is combined with the fairness model to enhance the mitigation process regarding execution time. The empirical evaluation shows that MAAT outperforms the existing approaches, thus mitigating the detected biases in less time. 

FairMASK \cite{peng_fairmask_2023} is a hybrid approach that exploits the explanation bias technique to infer possible biases before the training phase. In particular, the underpinning model is trained on non-protected attributes to use as the dependent feature in the classification task. SUbsequently, the approach performs the prediction phase by using a masking strategy to assess the overall performance. FairMASK has been compared with benchmarking tools, demonstrating that the adopted technique is more effective in mitigating group biases. 

Fan \etal propose a model-agnostic individual fairness testing approach, namely ExpGA, based on genetic algorithms \cite{10.1145/3510003.3510137}. The proposed strategy can handle black-box models by feeding the underlying model with the prediction probabilities, thus optimizing the fitness value. The approach is evaluated by considering \textit{i)} the overall performance, \textit{ii)} the execution time, and \textit{iii)} improvement through retraining.  

Conceived explicitly for NLP systems, ASTRAEA \cite{soremekun_astraea_2022} is a grammar-based instance-generation tool that identifies features causing fairness violations given an input model and mitigates them. To this end, it extracts sensitive attributes from the input grammar to cover individual and group fairness metrics.  

Perera \etal propose the \textit{Search-based Fairness Testing} (SBFT) tool to evaluate the individual fairness of ML regression systems \cite{perera_search-based_2022} based on the \textit{fairness degree} metric. The tool generates a set of unfair instances using a genetic algorithm approach. SBTF outperforms Aequitas and Themis in discovering individual unfairness.

Similar to \cite{10.1145/3377811.3380331}, NeuronFair \cite{9793943} exploits the adversarial strategy to generate individual discrimination instances (IDIs) and produce interpretable test cases for DNNs. The findings show that NeuronFair outperforms four baselines in terms of four different aspects, \ie effectiveness, efficiency, interpretability, and generalization, by considering seven different datasets.

Li \etal \cite{10.1145/3510003.3510091} propose a logistic-regression-based training data debugging (LTDD) strategy to remove group bias from training feature values. In this respect, the approach predicts the biased part of the features and removes them from the training samples to predict the final label. The proposed strategy outperforms state-of-the-art methodologies in terms of notable fairness indicators.

iRecSys2.0 is a fairness-aware in-process crowdworker recommendation system proposed by Wang \etal \cite{wang_context-_2022}. The proposed approach is optimized to overcome popularity bias by means of a multi-objective optimization-based re-ranking component. The authors evaluated their approach in terms of the effectiveness of the predictions and fairness.

The most relevant work to our approach is FairML \cite{10.1145/3550355.3552401}, an MDE-based approach specifically conceived to conceptualize fairness by relying on a tailored metamodel. The dedicated DSL covers the definition of bias and the actual assessment using predefined metrics. FairML eventually generates a YAML specification of the system that is compliant with the metamodel that the user can fine-tune.

\revision{Aequevox \cite{rajan_aequevox_2022} is a testing-based approach to test fairness in automatic speech recognition (ASR) systems. Based on a tailored definition of group bias in the ASR domain, the system first uses the metamorphic testing technique to locate possible bias in the preprocessed speech. Afterward, fault localization is employed to find unrepresented groups by employing Levenshtein distance. Aequevox has been evaluated on three different commercial ASR systems, showing that the approach can automatically identify group bias in different languages} 

d'Aloisio \etal propose MANILA \cite{manila2023}, a low-code tool to develop and execute fairness evaluation experiments by combining Software Product Lines and Extended Feature Models formalisms. 
After the feature selection phase, MANILA generates a Python implementation of the given experiment that evaluates each ML and fairness-enhancing method combination using the provided metrics. The authors evaluate their approach by replicating a case study, allowing the composition of different metrics.


\revision{Farify \cite{biswas_fairify_2023} assesses the fairness of neural networks model using satisfiability modulo theories (SMT) technique. Given a trained neural network and targeted fairness expressed as a SAT formula, \ie individual fairness, the approach applies input partitioning and sound pruning to identify neurons that are not activated. In addition, Fairify employs heuristic pruning to filter out neurons that can lead to bias, thus preserving fairness. The conducted experiment demonstrates that the approach is a light-weight solution for assessing fairness in neural networks.}

\revision {Similarly, DICE \cite{monjezi_information-theoretic_2023} is an automatic test-generation approach to detect individual bias in Deep Neural Networks (DNNs). As the first step, the approach generates test cases to identify the amount of discrimination for a given dataset. Then, the generated test have been used to locate neurons with a significant causal contribution to the discrimination. To assess the tool's effectiveness, DICE has been run on ten different datasets, showing that the approach can locate and mitigate individual bias.}


In summary, most reviewed tools primarily focus on applying pre-existing fairness definitions and metrics, ultimately conducting the final assessment within the social domain. Consequently, we recognize a compelling need to offer users a comprehensive and domain-agnostic framework that empowers them to define and evaluate their own bias and fairness criteria. \revision{We defer to Section \ref{sec:rq1} for a more comprehensive discussion, including a qualitative comparison with our approach in terms of the elicited features.}




\section{Proposed Approach}
\label{sec:proposedApproach}
In this section, we introduce \tool, a model-driven framework designed to conceptualize, design, implement, and execute the fairness assessment workflow illustrated in Figure \ref{fig:workflow}. Our approach can be characterized as a two-layered framework. At its core lies an abstract bias definition upon which multiple fairness analyses can be defined and built.

\begin{figure}[t!]
    \centering
    \includegraphics[width=\linewidth]{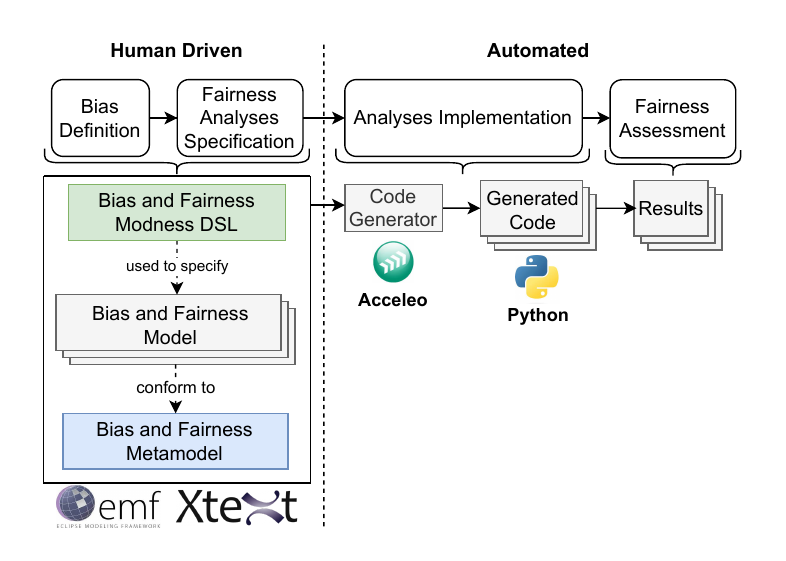}
    \caption{\revision{\tool high-level view.}}
    \label{fig:high_level_view}
\end{figure}

Figure \ref{fig:high_level_view} provides a high-level overview of \tool: round boxes represent the four primary steps of the fairness assessment workflow outlined in Section \ref{sec:motivation}. In contrast, square boxes depict the artifacts that are either developed or automatically generated. Adjacent to each artifact, we also indicate the technologies employed for its implementation. 
The fairness assessment workflow within \tool can be divided into two main phases: a \textit{Human Driven} phase, which involves user interaction (as described in Section \ref{sec:motivation}), and an \textit{Automated} phase, which operates without direct user involvement.

To initiate the fairness assessment process with \tool, the initial step involves defining bias by specifying the \textit{sensitive variables}, \textit{privileged and unprivileged groups}, and \textit{positive outcome}. \revision{Recalling the \textit{University} and \textit{TPL} use cases defined in Section \ref{sec:motivation}, the bias definition for \textit{University} could include \textit{gender} as the sensitive variable, \textit{men and women} as the privileged and unprivileged groups respectively, and \textit{positive admission} as the positive outcome. For the \textit{TPL} use case, a bias definition can comprise \textit{popularity} as the sensitive variable, \textit{popular and unpopular} libraries as the privileged and unprivileged groups, and \textit{recommendation} as the positive outcome. Note how these bias definitions are generic in this phase and unrelated to any specific dataset.} 

Subsequently, multiple analyses can be constructed based on a definition of bias. A \textit{fairness analysis} tailors a specific bias definition to a particular dataset with a defined scope and associated fairness metrics. These fairness metrics can be established in existing literature or custom-defined by the user. \revision{Note how the dataset may also include the predictions of an ML model in one of its columns, also allowing the fairness assessment of the model's predictions (similarly to other fairness toolkits like Aequitas or Themis \cite{saleiro_aequitas_2019,angell_themis_2018}).} Recalling our use cases, a fairness analysis for \textit{University} can be made of a dataset containing information about the gender of each student (mapped, for instance, in a column named \textit{sex})\footnote{\revision{Note how we refer to the original column names of the dataset related to this use case}}
and the admission outcome (for instance, in a column named \textit{admission}), the scope will be “\textit{equal probability for men and women to be admitted}," and the metric is \textit{Statistical Parity}. \rev{It is worth mentioning how this scope of the analysis belongs to what has been classified as \textit{separation} definitions of fairness (\eg Statistical Parity) \cite{castelnovo2022clarification}. Another possible scope could be “\textit{to have, for each group, the same probability of a positive outcome while allowing differences based on relevant features}." This scope belongs to what has been defined as \textit{independence} fairness definitions \cite{castelnovo2022clarification}, and possible metrics compliant with this scope could be \textit{Equal Opportunity} or \textit{Average Odds} \cite{hardt_equality_2016}. This aspect highlights the layered structure of MODNESS, where from a high-level definition of bias multiple analyses can be depicted.}

Concerning the \textit{TPL} use case, a fairness analysis can comprise a dataset with the popularity of each library (for instance, in a column named \textit{frequency}) and a recommendation score (for instance, in a column named \textit{recommendation}). The recommendation score is eventually used by the recommender system to identify the items more suited for a recommendation. For instance, the system may recommend only the items with a score higher than 80\% of all other libraries. The scope of the analysis will be that \textit{each library must be recommended despite its popularity}, and the metric adopted is the custom \textit{coverage} metric from the recommender systems domain \cite{robillard2009recommendation}. The traditional \textit{coverage} metric measures the number of items being recommended over the total amount of items \cite{NGUYEN2019110460}. This variation measures the amount of \textit{unpopular} libraries recommended over the whole recommendations. It is defined as:
$|L_{unpop}|/|L|$
where $|L_{unpop}|$ is the number of \textit{unpopular} (\ie \textit{non-frequent}) libraries recommended and $|L|$ is the whole set of recommendations \cite{nguyen2023dealing}. The closer this metric is to 1, the more the system is free from popularity bias.


These specification steps are implemented in \tool as a model-driven approach, utilizing the EMF ecosystem \cite{steinberg2008emf}. In this phase, the output is a model (referred to as the \textit{Bias and Fairness Model} in Figure \ref{fig:high_level_view}) that includes both the bias and the corresponding fairness analysis definitions. This model adheres to the \textit{Bias and Fairness Metamodel}, which serves as the foundational structure of \tool.

Starting from a model describing the bias and the related fairness analyses definitions, \tool automatically generates an implementation of them through a code generator based on Acceleo.\footnote{\url{https://www.eclipse.org/acceleo/}} In particular, \tool generates a Python code that automatically checks the fairness of a given dataset using the information provided in the fairness analysis specification. 
The analyses implementation and the fairness assessment steps comprise the automated phase of \tool and do not require direct human intervention.


\revision{To support the specification of bias and fairness models, we developed a domain-specific language and its corresponding environment utilizing Xtext technology.\footnote{\url{https://eclipse.dev/Xtext/}} This way, users can rely on a textual concrete syntax to specify all the concepts needed in the traditional fairness workflow. In the forthcoming section, we present the textual \tool specification for two explanatory use cases, \ie \textit{University} and \textit{TPL}, while presenting the \tool metamodel.
The Xtext grammar of the \tool DSL is available online at \cite{Daloisio_MODNESS_2024}.}

In the following, we detail \tool by first describing the implemented metamodel in Section \ref{sec:metamodel} and, next, describing the code generation and fairness assessment processes in Section \ref{sec:generation}.\rev{The source code of \tool is available in our replication package \cite{Daloisio_MODNESS_2024}}.

\subsection{Bias and Fairness Metamodel}\label{sec:metamodel}

\begin{figure}[ht!]
    \centering
    \includegraphics[width=\linewidth]{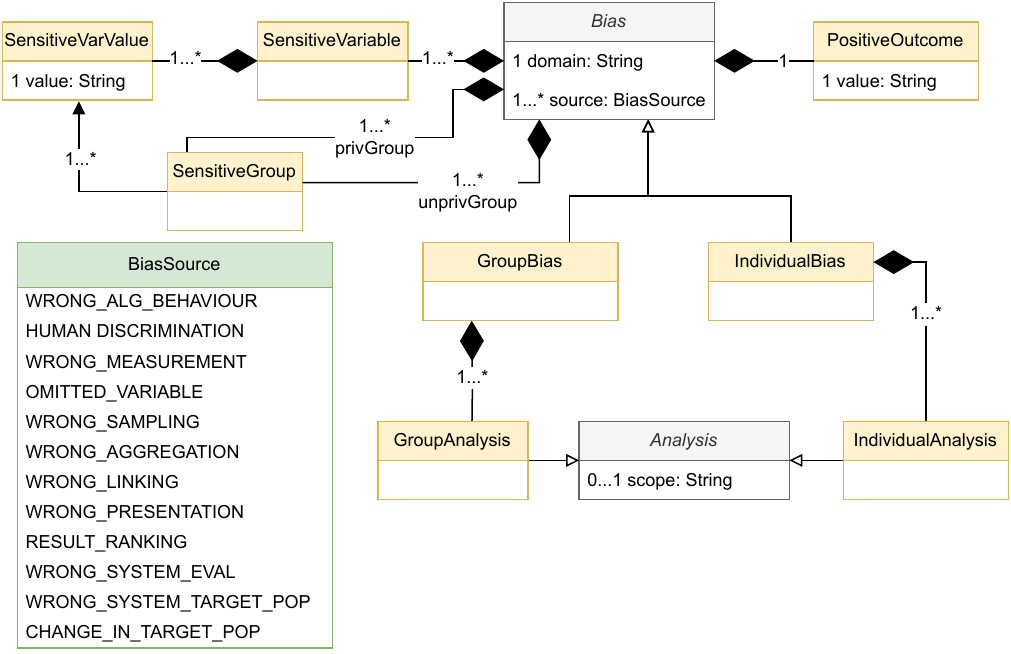}
    \caption{\revision{Bias Definition.}}
    \label{fig:bias_metamodel}
\end{figure}

\begin{figure}[b!]
    \centering
    \includegraphics[width=\linewidth]{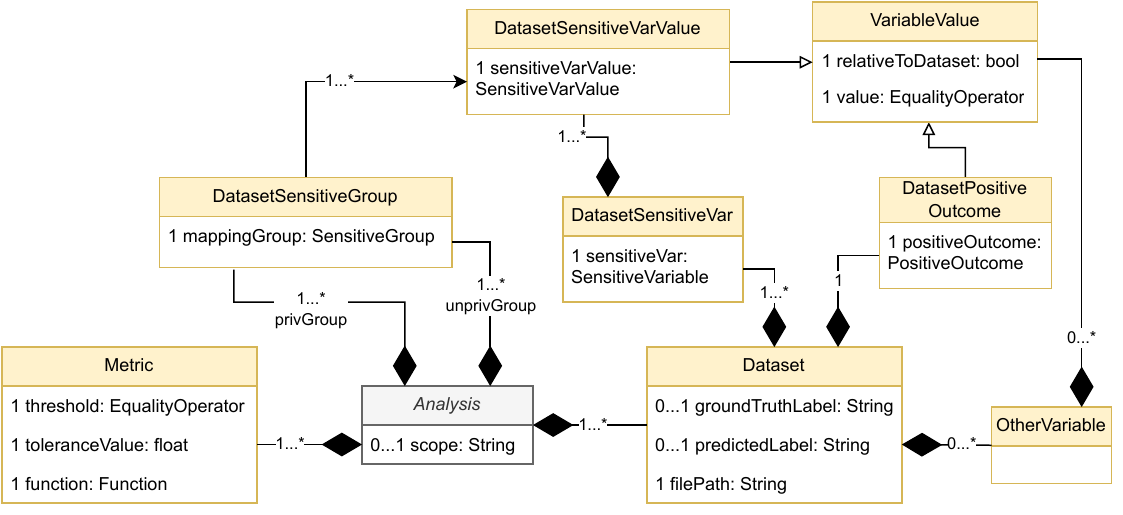}
    \caption{Fairness Analysis.}
    \label{fig:fairness_analysis}
\end{figure}

\begin{figure}[ht!]
    \centering
    \includegraphics[width=\linewidth]{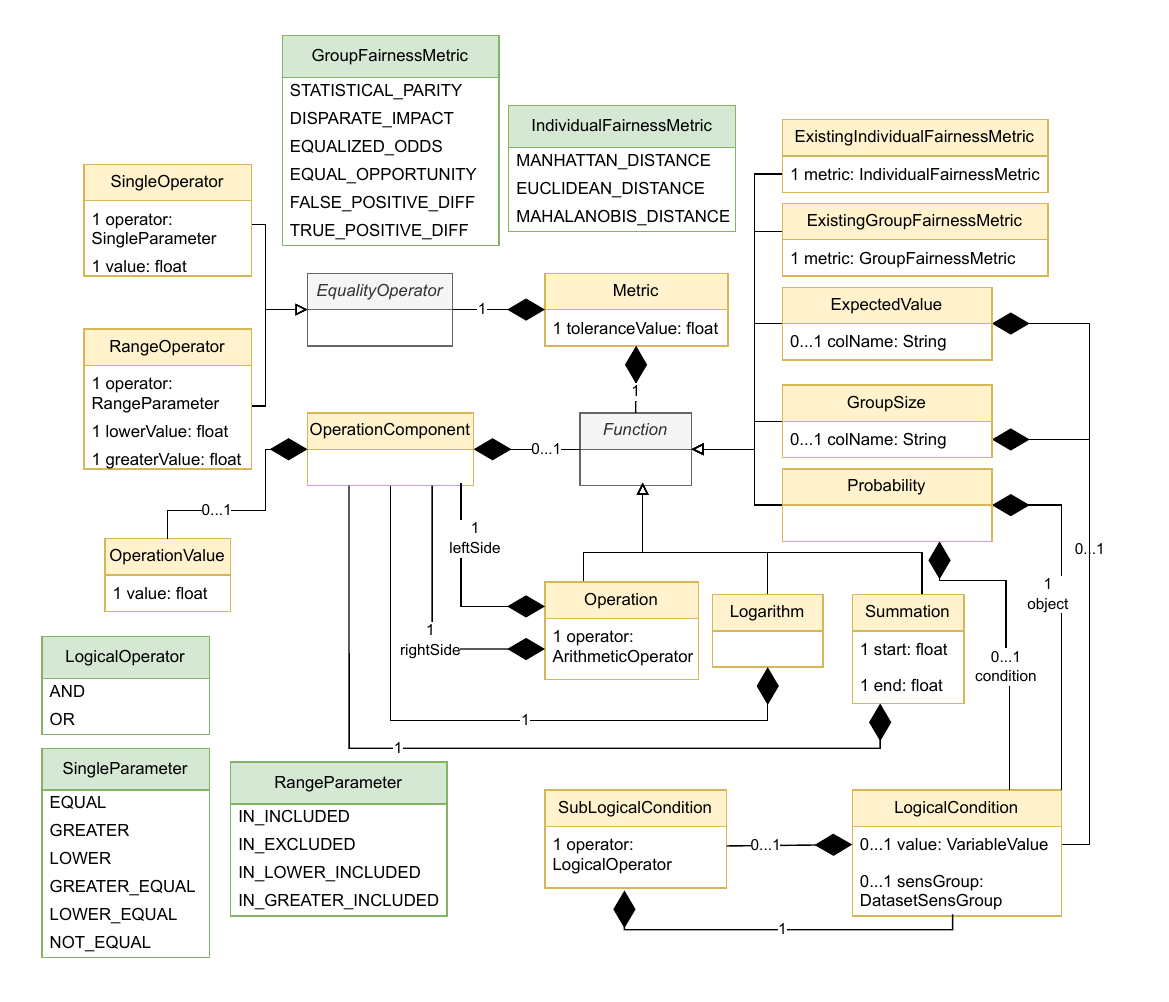}
    \caption{Metric Definition.}
    \label{fig:metric_def}
\end{figure}

The bias and fairness metamodel is the foundation of \tool. It allows users to define the bias related to a particular domain and to specify fairness analyses starting from that definition. Hence, the metamodel can be divided into three related packages providing the modeling constructs for bias definitions (see Fig. \ref{fig:bias_metamodel}), fairness analyses specification (see Fig. \ref{fig:fairness_analysis}), and for the definition of metrics (see Fig. \ref{fig:metric_def}).

\revision{\subsubsection{Bias Definition}}

The root of the metamodel in Figure \ref{fig:bias_metamodel} is the abstract class \texttt{Bias} representing the concept of discrimination at a higher level of abstraction. Bias has a domain and one or more sources (\ie what generated bias). The possible sources of bias have been selected from \cite{mehrabi_survey_2021} \eg \textit{human discrimination}, \textit{wrong sampling of data}, among others \revision{and have been listed in the dedicated \texttt{BiasSource} enumeration}. Then, bias is composed of a \texttt{PositiveOutcome} and one or more \texttt{SensitiveVariable} instances, which have one or more \texttt{SensitiveVarValue} each. Finally, both the privileged and unprivileged groups must be specified in defining bias. The \texttt{SensitiveGroup} metaclass models these groups. In particular, each \texttt{SensitiveGroup} is identified by one or more \texttt{SensitiveVarValue}. The \texttt{Bias} metaclass is then specialized by two sub-metaclasses representing \texttt{GroupBias} and \texttt{IndividualBias}. Both group and individual biases are composed of one or more \texttt{Analysis} with a particular \texttt{Scope}. In particular, \texttt{GroupBias} has one or more \texttt{GroupAnalysis}, while \texttt{IndividualBias} has one or more \texttt{IndividualAnalysis}. 

\begin{lstlisting}[style=DSLStyle,caption=\revision{Bias definition example for the \textit{University} use case.},label=lst:bias-university]
GroupBias "university"{
	Definition: {
		domain: "education";
		source: WRONG_SAMPLING;
		sensitiveVariables: {
			SensitiveVariable{
				name: "gender";
				values: "male","female";
			}
		};
		positiveOutcome: "positive admission";
		unprivilegedGroup: {
			SensitiveGroup{
				name: "women";
				sensitiveValue: "gender.female";
			};
		};
		privilegedGroup: {
			SensitiveGroup{
				name: "men";
				sensitiveValue: "gender.male";
			};
		};	
	};
\end{lstlisting}


\revision{Listing \ref{lst:bias-university} shows an example of \tool bias definition related to the \textit{University} use case. Since this use case is about group bias, the model's root is an instance of the \texttt{GroupBias} metaclass. Next, the first portion of the model consists of a \textit{ definition}, which specifies all the high-level components of a bias definition. For this use-case, the domain can be \textit{education}, and the source can be \textit{wrong sampling} (\eg wrong data have been used to train the ML model). Recalling the general workflow described in Section \ref{sec:motivation}, to give a high-level definition of group bias, we must provide the sensitive variables, the positive outcome, and the privileged and unprivileged groups. For this scenario, we have only one sensitive variable representing the \textit{gender}, which has two values, \ie \textit{male} and \textit{female}.\footnote{\revision{Note how the values \textit{male} and \textit{female} are instances of the \texttt{SensitiveVarValue} metaclass. However, in the DSL, they are represented as \texttt{values} attributes of the \texttt{SensitiveVariable} metaclass so as not to burden the overall syntax.}}
Next, the \textit{positive outcome} is represented by a \textit{positive admission}. Finally, the \textit{unprivileged group} is \textit{women} and has a reference to the \textit{female} sensitive value (indicating that this group is identified by that specific value of the sensitive variable \textit{gender}). On the contrary, the \textit{privileged group} is \textit{men} and has a reference to the \textit{male} sensitive value.}

\begin{lstlisting}[style=DSLStyle,caption=\revision{Bias definition example for the \textit{TPL} use case.},label=lst:bias-tpl]
GroupBias "TPL"{
	definition: {
		domain: "recommender systems";
		source: WRONG_ALGORITHM_BEHAVIOUR;
		sensitiveVariables: {
			SensitiveVariable{
				name: "popularity";
				values: "popular","unpopular";
			}
		};
		positiveOutcome: "recommendation";
		unprivilegedGroup: {
			SensitiveGroup{
				name: "unpopular libraries";
				sensitiveValue: "popularity.unpopular";
			};
		};
		privilegedGroup: {
			SensitiveGroup{
				name: "popular libraries";
				sensitiveValue: "popularity.popular";
			};
		};	
	};	
\end{lstlisting}

\revision{Listing \ref{lst:bias-tpl} reports instead an example of bias definition for the \textit{TPL} use case. Note how the main components of a bias definition are always the same regardless of the domain (\ie \textit{group} or \textit{individual} bias, \textit{sensitive variables}, \textit{positive outcome}, and \textit{sensitive groups} if the definition is a group bias). The root of the model is a \texttt{GroupBias} class where the domain is \textit{recommender systems}, and the source of bias could be \textit{wrong algorithm behaviour}. The \textit{sensitive variable}, in this case, is represented by \textit{popularity} its possible values are \textit{popular} and \textit{unpopular}. The positive outcome is a \textit{recommendation} from the system. Finally, the \textit{unprivileged} group is represented by unpopular libraries, whereas the \textit{privileged} group is represented by popular libraries.} 

\revision{It is worth noticing how, being high-level and not related to a specific dataset or analysis, this portion of the model can be defined by the domain and legal experts only, without assistance from data scientists.}

\revision{\subsubsection{Fairness Analysis}}

Figure \ref{fig:fairness_analysis} depicts the metamodel portion dedicated to the fairness analysis specification, represented by the \texttt{Analysis} abstract metaclass. An analysis may have a scope, \ie a textual description of the analysis, and is composed by one or more \texttt{Dataset}s. A \texttt{Dataset} has an attribute to specify the name of the column containing the \texttt{groundTruthLabel} (if any), an attribute to specify the name of the column containing the \texttt{predictedLabel} (if available), and an attribute to specify its \texttt{filePath}. Then, a mapping of each general concept defined in the bias definition must be identified in the dataset. In particular, a \texttt{Dataset} is composed of one \texttt{DatasetPositiveOutcome} metaclass mapping the value of the positive outcome in the dataset, one or more \texttt{DatasetSensitiveVar} metaclasses (which are in turn composed of one or more \texttt{DatasetSensitiveVarValue} metaclasses) mapping the sensitive variables, and, if needed, one or more \texttt{OtherVariable} metaclasses representing other values encoded in the dataset. Then the sensitive groups must also be mapped in the dataset through the \texttt{DatasetSensitiveGroup} metaclass. All the metaclasses representing values extend a \texttt{VariableValue} metaclass. It is worth noting that all the values can be absolute or relative to the dataset. Finally, an analysis comprises one or more \texttt{Metric}.


\begin{lstlisting}[style=DSLStyle,caption=\revision{Fairness analysis example for the \textit{University} use case.},label=lst:university_analysis]
analysis: {
    GroupAnalysis{
        scope: "all people must have 
                same admission 
                probability despite gender";
        dataset: {
            Dataset {
                id: 'admissions';
                predictedLabelName: 'admitted';
                filePath: 'admissions.csv';
                positiveOutcome: {
                    id: "admission";
                    mappingOutcome: "positive admission";
                    value: {
                        operator: EQUAL;
                        value: 1.0;
                    };
                };
            datasetSensitiveVariable: {
                DatasetSensitiveVariable{
                name: "sex";
                mappingSensitiveVariable: gender;
                values: {
                    SensitiveVariableValue{
                        id: "female";
                        mappingValue: "gender.female";
                        value: {
                            operator: EQUAL;
                            value: 0.0;	
                        };
                    },
                    SensitiveVariableValue{
                        id: "male";		
                        mappingValue: "gender.male";
                        value: {
                            operator: EQUAL;
                            value: 1.0;
                        };
                    }
                    }
                }
            }; 
        }
    };
    datasetUnprivilegedGroup: {
            id: 'women';
            mappingGroup: women;
            sensitiveVariables: 
            ("admissions.sex.female");
    };
    datasetPrivilegedGroup: {
        id: 'men';
        mappingGroup: men;
        sensitiveVariables: 
        ("admissions.sex.male");  
    };	
\end{lstlisting}

\revision{Listing \ref{lst:university_analysis} shows the fragment of the model related to the analysis definition for the \textit{University} scenario. We recall that an analysis implements a high-level bias definition and maps each abstract component into a real feature of a dataset. In this example, since we start from a \texttt{GroupBias} definition, the analysis is modelled as an instance of the \texttt{GroupAnalysis} metaclass. The scope of the analysis is that \textit{``all people must have the same admission probability despite their gender"}. 
The analysis comprises a \texttt{Dataset} class, which models the dataset that will be used in the analysis. In particular, the model's predictions (\ie admission of students) are stored in a column named \textit{admitted} (see line 9 of Listing \ref{lst:university_analysis}). 
As specified in the bias definition, a positive outcome for this use case is represented by a \textit{positive admission}, which is encoded with a value of $1.0$ in the \textit{admitted} column of the dataset. This information is represented in the model by the \texttt{positiveOutcome} attribute of the Dataset class, which says that the \textit{positive admission} positive outcome is represented with a \textit{value} equal to $1.0$ (lines from 11 to 18).}

\revision{Next, we need to associate each \textit{sensitive variable} with specific columns and values in the dataset. More specifically, the model defines the \texttt{"sex"} column as representing the \textit{gender} sensitive variable, where \textit{female} is coded as a value of $0.0$, and \textit{male} is coded as a value of $1.0$ (referenced in lines 19 to 44).}

\revision{Furthermore, the model identifies the unprivileged group as \textit{women} within the dataset, which corresponds to instances with a value of $0.0$ in the \texttt{"sex"} column (this concept is captured within the model through a linkage to the \texttt{SensitiveVariableValue} class, marked by the ID \texttt{"admissions.sex.female"}). Similarly, the privileged group is represented by instances that have a value of $1.0$ in the dataset (covered in lines 45 to 56).}


\begin{lstlisting}[style=DSLStyle,caption=\revision{Fairness analysis example for the \textit{TPL} use case.},label=lst:tpl_analysis]
analysis: {
    GroupAnalysis{
        scope: "relevant libraries must 
                be recommended despite 
                their popularity";
        dataset: {
            Dataset {
                id: 'recommendations';
                predictedLabelName: 'ranking';
                filePath: 'recommendations.csv';
                positiveOutcome: {
                    id: "high-ranking";
                    mappingOutcome: recommendation;
                    value: {
                        operator: GREATER_EQUAL;
                        value: 0.8;	
                    };
                    relativeToDatasetSize;	
                };
            datasetSensitiveVariable: {
                DatasetSensitiveVariable{
                    name: "frequency";
                    mappingSensitiveVariable: 
                    popularity;
                    values: {
                        SensitiveVariableValue{
                            id: "non-frequent";
                            mappingValue: 
                            "popularity.unpopular";
                            value: {
                                operator: MINOR_EQUAL;
                                value: 0.8;	
                            };
                            relativeToDatasetSize;
                        },
                        SensitiveVariableValue{
                            id: "frequent";
                            mappingValue: 
                            "popularity.popular";
                            value: {
                                operator: GREATER;
                                value: 0.8;
                            };
                            relativeToDatasetSize;
                        }
                    }
                }
            };	
        }
    };
    datasetUnprivilegedGroup: {
            id: 'non-frequent libraries';
            mappingGroup: "unpopular libraries";
            sensitiveVariables: ("recommendations.frequency.non-frequent");
    };
    datasetPrivilegedGroup: {
        id: 'frequent libraries';
        mappingGroup: "popular libraries";
        sensitiveVariables: ("recommendations.frequency.frequent");
    };
\end{lstlisting}

\revision{Listing \ref{lst:tpl_analysis} shows the analysis definition for the \textit{TPL} use case. Similar to the \textit{University} scenario, this analysis maps each component of the bias definition into concrete features of the dataset under analysis. We start with an instance of the \texttt{GroupAnalysis} metaclass. An instance of the \texttt{Dataset} metaclass models the dataset used for the analysis, with the predicted rank stored in the \textit{ranking} column. This information is thus reported as an attribute of the \texttt{Dataset} class (line 9). Additionally, based on inputs from the domain expert, we know that the system recommends a library if its ranking exceeds 80\% of the predicted ranks (\ie if the libraries are ordered in descending rank order, only the top 20\% are recommended) \cite{nguyen2023dealing}. This can be modelled in \tool by specifying in the \texttt{positiveOutcome} attribute that a \textit{recommendation} positive outcome is encoded with a value greater than or equal to $0.8$ for the \textit{ranking} class. The \texttt{relativeToDatasetSize} keyword indicates that we are using relative values (\ie percentage) rather than absolute ones (lines 11-19).}

\revision{Similarly, assume that the domain expert specifies that a library is \textit{popular} if it appears in many projects \cite{nguyen2023dealing}. In particular, the dataset has a column named \textit{frequency} containing the number of projects in which it appears, and a library is \textit{popular} if its frequency is higher than 80\% of all libraries. Hence, first, an instance of the \texttt{DatasetSensitiveVariable} metaclass maps the \textit{popularity} sensitive variable to the \textit{frequency} column in the dataset (lines 21-24). Next, as done for the ranking, the model maps \textit{popular} libraries to values greater than $0.8$ using the \texttt{relativeToDatasetSize} keyword and \textit{unpopular} libraries to values lower or equal to 0.8 of the entire dataset (lines 25-45). Finally, the model reports how the unprivileged group is identified in the dataset with the \textit{non-frequent} sensitive variable value, while the privileged group is identified in the dataset with the \textit{frequent} sensitive variable value (lines 51-60).}


\revision{\subsubsection{Metric Definition}\label{sec:metric_def}}

Figure \ref{fig:metric_def} reports the portion of the metamodel dedicated to the metric definition. A \texttt{Metric} is composed of an \texttt{EqualityOperator} representing the threshold and has an attribute representing the \texttt{toleranceValue} (\ie the level of bias tolerated by the system). The \\ \texttt{EqualityOperator} can be a \texttt{SingleOperator} (\eg "$=0$" or "$\le 1$") or a \texttt{RangeOperator} (\eg "the metric must be $< 1$ and $>0$"). Next, a \texttt{Metric} has a \texttt{Function} representing the actual metric implementation. Currently, the following functions are available: \texttt{Operation} (representing a generic arithmetical operation), \texttt{Logarithm}, \texttt{Summation}, \texttt{ExpectedValue}, \texttt{GroupSize}, and \texttt{Probability}. A metric can also be based on \texttt{ExistingGroupFairnessMetric} and \texttt{Existing\-IndividualFairnessMetric}. Such metaclasses represent the set of fairness metrics known in the literature and already implemented for group and individual fairness definitions, respectively.\footnote{To select the set of metrics, we referred to the ones implemented in the AIF360 library \cite{bellamy_ai_2019}.} For each function, we included a set of attributes needed for their implementation. In the future, new functions can be added by extending the \texttt{Function} metaclass. 


\begin{lstlisting}[style=DSLStyle,caption=\revision{Metric definition example for the \textit{University} use case.},label=lst:univ_metric]
metric: {
        Metric{
            name: "StatisticalParity";
            toleranceValue: 0.2;
            function: ExistingGroupFairnessMetric { 
                metric: STATISTICAL_PARITY; 
            };
            threshold: {
                operator: EQUAL;
                value: 0.0;
            };
        };
};
\end{lstlisting}

\revision{Listing \ref{lst:univ_metric} shows the fragment of the model devoted to the metric definition for the \textit{University} use case. In this scenario, based on the scope of the analysis, the data scientist suggests using the \textit{Statistical Parity (SP)} metric to assess fairness. SP is a widely adopted metric in the fairness literature that measures the probability of an item receiving a positive prediction, whether it is in the privileged group or not \cite{dwork_fairness_2012}. A value of 0 means fairness. This metric is included among the possible values for the \texttt{ExistingGroupFairnessMetric} metaclass. Hence, to model this information in \tool, a user first has to define an instance of the \texttt{Metric} metaclass and set a tolerance value, \eg $0.2$ \cite{feldman_certifying_2015}. Next, they have just to define an instance of the \texttt{ExistingGroupFairnessMetric} metaclass and set its value to \texttt{STATISTICAL\_PARITY}. Finally, this metric's optimal value is reported as equal to $0.0$.}


\begin{lstlisting}[style=DSLStyle,caption=\revision{Metric definition example for the \textit{TPL} use case.},label=lst:tpl_metric]
Metric{
        name: "coverage";
        toleranceValue: 0.2;
        function: Operation{
            arithmeticOperator: RATIO;
            leftSide: {
                function: GroupSize{
                    groupCondition: {
                        sensitiveGroup: "non-frequent libraries"
                        AND value:"recommendations.high-ranking"
                    };
                };
            };
            rightSide: {
                function: GroupSize{
                    groupCondition: {
                        value: "recommendations.high-ranking"
                    };
                };
            };
        }
        threshold: {
            operator: EQUAL;
            value: 1.0;
        };	
    };
\end{lstlisting}

\revision{
Listing \ref{lst:tpl_metric} reports instead an implementation of a metric for the \textit{TPL} use case. In this case, the data scientist suggests using a custom metric to assess the amount of bias in the system. In particular, they suggest using an adaptation of the \textit{coverage} metric from the recommender systems domain to measure popularity bias \cite{robillard2009recommendation}. Recall, from the definition given in Section \ref{sec:proposedApproach}, that this metric is defined as the ratio of \textit{unpopular} items recommended over the whole recommendations, \ie $|L_{unpop}|/|L|$. 
Since this metric is not a traditional metric from the fairness literature, it is not included among the possible values for the \texttt{GroupFairnessMetric} enumeration. However, it can be modelled in \tool as follows: first, create an instance of the \texttt{Metric} metaclass, which will contain the custom metric and set its tolerance value, for instance, $0.2$ (lines 2-3 in Listing \ref{lst:tpl_metric}). Next, recalling the definition given above, this metric is a ratio between two values. Hence, we create an instance of the \texttt{Operation} metaclass with the \texttt{arithmeticOperator} attribute set to \texttt{RATIO} (line 5). Next, we have to model the left and right sides of the ratio (\ie its numerator and its denominator). 
The numerator is the number of \textit{non-frequent} libraries being recommended.
This information can be modelled in \tool by first creating an instance of the \texttt{GroupSize} metaclass, which represents a function to count the number of items in a group (line 7). Next, we have to define the set of items that have to be counted by the \texttt{GroupSize} function (\ie the \textit{non-frequent} items being recommended). So, we define a set of boolean conditions that can be used to select a set of items from the dataset. In this case, we have two logical conditions connected with an \texttt{AND}. The first logical condition selects items from the \textit{non-frequent} sensitive group, while the second logical condition selects items having the \textit{high-ranking} positive outcome (lines 9-10). The denominator is instead the number of items having a \textit{high-ranking}. This information can be modelled similarly to the numerator by creating an instance of \texttt{GroupSize} metaclass, this time filtering only items with a \textit{high-ranking} (lines 15-19). Finally, it is reported that the threshold for this metric is equal to 1.0 (lines 22-25).
}

\subsection{Code generation and fairness assessment}\label{sec:generation}

After defining fairness and its corresponding metrics, \tool generates the fairness assessment implementation using a source code generator developed with Acceleo. This generator exploits specific templates to create static and dynamic parts of the target code by incorporating queries on the source models. Acceleo templates leverage a defined syntax to specify conditions or iterations over elements in the input models. Listing \ref{lst:template} depicts an excerpt of the developed Acceleo template.\footnote{\revision{The developed Acceleo-based code generator is available on our replication package \cite{Daloisio_MODNESS_2024}}}

\begin{lstlisting}[label=lst:template,caption=Fragment of an explanatory \tool Acceleo template.,style=JavaStyle]
file_path = '[biasModel.dataset.filePath/]'
predicted_label_name = '[biasModel.dataset.predictedLabelName/]'
ground_truth_label_name = '[biasModel.dataset.groundTruthLabelName/]'
...
[if biasModel.metric -> size()>0]
[for (metric: Metric | biasModel.metric)]
[if (metric.operator.oclIsTypeOf(SingleOperator))]
threshold = [metric.operator.oclAsType(SingleOperator).value/]
[/if]
[/for]
[/if]
\end{lstlisting}

To support the generation phase, we rely on Pandas \cite{mckinney-proc-scipy-2010} and AI360 Python \cite{bellamy_ai_2019} libraries to preprocess and support fairness analysis, respectively. In particular, \tool exploits three different Acceleo templates developed to support each phase of the process, \ie bias definition, fairness analysis specification, and metric definition. To cover the first phase, the generated code supports the high-level specification of the analyzed fairness scenario, including the sensitive variables, the expected positive outcome, and the parameters needed to feed the selected metric. In such a way, the configuration is compliant with the user-defined specification. Afterwards, the fairness analysis phase could be conducted by means of a set of predefined implementations of the state-of-the-art fairness metrics (taken from the AIF360 library \cite{bellamy_ai_2019}). Alternatively, \tool can generate operator specifications that define and compose different metrics. 

\begin{lstlisting}[label=lst:university,caption=\revision{Generated code for the \textit{University} use case.},style=PythonStyle]
from fairnessMetric import FairnessMetric
import pandas as pd

# INPUT DATA
file_path = "data/admissions.csv"
predicted_label_name = "admission"
data = pd.read_csv(file_path)
indexes = ["sex"]
dataset_unprivileged_group = {"sex": 0}
dataset_privileged_group = {"sex": 1}

# PARAMETERS
dataset_positive_outcome = 1
threshold = 0.0
tolerance_value = 0.2

# FAIRNESS ASSESSMENT
metrics = FairnessMetric(data,dataset_unprivileged_group,dataset_privileged_group,ground_truth_label_name,predicted_label_name,dataset_positive_outcome)
print(metrics.statistical_parity_difference())
if abs(metrics.statistical_parity_difference()) > (threshold + tolerance_value):
    print("Biased")
else:    
    print("Fair")
\end{lstlisting}

Listing \ref{lst:university} reports the code generated by \tool for the \textit{University} use case. All the metrics and operations defined in the metamodel (\ie all the metaclasses extending the \texttt{Function} metaclass in Fig. \ref{fig:metric_def}) have been implemented as functions in the \texttt{FairnessMetric} Python class, which is imported at the beginning of the script. In addition, the \texttt{pandas} Python library is imported to read and process the dataset. Next, lines from \texttt{4} to \texttt{15} define a set of variables implementing attributes specified in the model during the \textit{fairness analysis} definition phase (\ie file path, predicted label name, privileged and unprivileged groups, positive outcome, threshold, and tolerance value). 
Finally, the \texttt{FairnessMetric} class is instantiated on line \texttt{18}. As said above, this class provides both implementations of existing fairness metrics and operations to create new ones. Since, in this use case, we are using a metric already defined among the possible values of the \texttt{GroupFairnessMetric} enumeration (\ie \textit{statistical parity}),  the generated code simply calls a method from the \texttt{FairnessMetric} class implementing it (line \texttt{20}). Finally, lines from \texttt{21} to \texttt{24} check if the value returned by the metric is within the defined threshold. If so, a \texttt{"Fair"} message is printed, \texttt{"Biased"} otherwise.

\begin{lstlisting}[style=PythonStyle,caption=\revision{Generated code for the \textit{TPL} use case.},label=lst:tpl]
from FairnessMetric import FairnessMetric, binarize
import pandas as pd
from operators import SingleOperator

# INPUT DATA
file_path = "data/popbias.csv"
predicted_label_name = "ranking"
data = pd.read_csv(file_path)

# PREPROCESSING
operator_value = 0.8
operator = SingleOperator.GREATER_EQUAL
binarize(data, "frequency", operator, operator_value, True)
operator_value = 0.8
operator = SingleOperator.GREATER_EQUAL
binarize(data, "ranking", operator, operator_value, True)

# PARAMETERS
dataset_unprivileged_group = {"frequency": 0}
dataset_privileged_group = {"frequency": 1}
dataset_positive_outcome = 1
threshold = 1.0
tolerance_value = 0.2

# FAIRNESS ASSESSMENT
metrics = FairnessMetric(
    data,
    dataset_unprivileged_group,
    dataset_privileged_group,
    ground_truth_label_name,
    predicted_label_name,
    dataset_positive_outcome,
)
coverage = metrics.group_size("frequency == 0 and ranking == 1") / metrics.group_size("ranking == 1")
print(coverage)
if abs(coverage) < threshold + tolerance_value:
    print("Biased")
else:
    print("Fair")
\end{lstlisting}

Listing \ref{lst:tpl} reports instead the generated code for the \textit{TPL} use case. The code follows the same structure of Listing \ref{lst:university}, with some additional changes. The first difference is shown on lines from \texttt{12} to \texttt{17}. In particular, differently from the \textit{University} use case where all the values were binary, here both \textit{frequency} and \textit{ranking} columns contain continuous values. Hence, based on the definition of \textit{positive label} and \textit{sensitive groups} specified in the model, those lines of code map values greater or equal than 0.8 to 1 and values lower than 0.8 to 0. This binary mapping is needed because all the fairness metrics available are defined on binary data \cite{mehrabi_survey_2021,verma_fairness_2018}. The second main difference is about the adopted metric. As stated in Section \ref{sec:metric_def}, in this use case, we use a custom fairness metric named \texttt{coverage}. Since this is a custom metric, differently from the \textit{University} use case, it is not directly defined as a function in the \texttt{FairnessMetric} class. Instead, line \texttt{35} shows how this metric is implemented in the code. In particular, it is implemented as the ratio between the values returned by two calls of the \texttt{group\_size} function. The input of the first function (\ie the numerator) is a string selecting libraries with \texttt{frequency} equal to 0 (\ie unpopular libraries) and \texttt{ranking} equal to 1 (\ie recommended libraries). The input of the function in the denominator is instead a string selecting only recommended libraries (\ie \texttt{ranking} equal to 1).

\begin{lstlisting}[style=PythonStyle,caption=\rev{Portion of the \texttt{FairnessMetric} class.},label=lst:fairmetr]
from aif360.metrics import ClassificationMetric
from aif360.datasets import BinaryLabelDataset
import pandas as pd
import math

class FairnessMetric(ClassificationMetric):
    def __init__(
            self,
            df: pd.DataFrame,
            unprivileged_groups: dict,
            privileged_group: dict,
            true_label_name: str,
            predicted_label_name: str,
            positive_value: int
        ):
            ...
            super().__init__(
                self.dataset_true,
                self.dataset_pred,
                unprivileged_groups=[unprivileged_groups],
                privileged_groups=[privileged_group])
    
     def probability(
            self, 
            object: str, 
            condition: str = ""
    ) -> float:
            probability = self.df.query(object).shape[0] / self.df.shape[0]
            if condition == "":
                return probability
            else:
                return (
                    self.df.query(condition + " and " + object).shape[0] 
                    / self.df.shape[0]
                ) / probability
    ...
\end{lstlisting}

\rev{Finally, Listing \ref{lst:fairmetr} reports a portion of the \texttt{FairnessMetric} class implementation. As shown in line \texttt{6}, this class extends the \texttt{ClassificationMetric} class from the \texttt{aif360} library. Thus, it inherits all the standard fairness metric implementations (\ie the ones defined in the \texttt{GroupFairnessMetric} and \texttt{IndividualFairnessMetric} enumerations reported in Figure \ref{fig:metric_def}) from this library. In addition, this class provides implementations of the functions reported in Figure \ref{fig:metric_def} to define custom metrics. For instance, in Listing \ref{lst:fairmetr} it is reported the implementation of the \texttt{probability} function.}


\section{Evaluation}
\label{sec:discussion}
\revision{In this section, we present the evaluation of the proposed approach by referring to the RQs introduced in Section \ref{sec:introduction}. In particular, we rigorously assess the \textit{expressiveness} and \textit{correctness} of \tool, demonstrating how it effectively addresses the limitations of contemporary state-of-the-art methods in assessing fairness.} 

In the following, we first present the set of use cases examined to answer the RQs. Then, we address and answer each RQ. 

\subsection{Examined use cases} \label{sec:scenarios}

By carefully analyzing the approaches described in Table \ref{tab:toolComparison}, we have identified pertinent use cases utilized across the literature to evaluate fairness within various domains. Given the vast array of potential scenarios, it is impractical to examine all of them within this section. Therefore, we focus on widely embraced case studies, specifically those covered by at least three distinct approaches. 
\revision{Moreover, to emphasize the expressiveness of \tool}, we introduce two additional case studies into our comparative analysis.
These case studies pertain to the evaluation of popularity bias in recommender systems encompassing a curated dataset of third-party Java libraries and a curated dataset of Arduino hardware and software components, respectively. \revision{Table \ref{tab:useCases} reports a short description of these use cases as well as references to the approaches that have addressed them. Note how two of these use cases have been used as running examples throughout this paper(\ie \textit{University} and \textit{TPL}) and are highlighted in the table. In the following, we provide detailed descriptions of the other use cases and their associated datasets.}

\revision{It is important to notice how, in this evaluation, we follow the related literature to identify the \textit{sensitive group(s)} and the \textit{positive outcome}. In a normal use case, the sensitive groups are instead identified based on the outcome of the fairness assessment, starting from a set of possible sensitive variables (\eg variables that are protected by regulations \cite{feldman_certifying_2015}) and an outcome considered \textit{positive} for the specific use case.}


\smallskip
\noindent \ding{228} \textbf{ProPublica Recidivism (COMPAS)  \cite{angwin_machine_2016} -}    
The \textit{Correctional Offender Management Profiling for Alternative Sanctions} (COMPAS) was an ML system used by judges in the US to predict if a condemned person would have been a repeating offender in the two years after their release. An investigation of this software showed that this system had a bias against \textit{non-white women}. In this case, the sensitive variables are \textit{race} and \textit{gender}, the favourable outcome is \textit{non-recidiv}, and the unprivileged group is \revision{\textit{non-white men}}.

\smallskip
\noindent \ding{228} \textbf{Adult Census Income (ADULT) \cite{kohavi_scaling_nodate} -}
This use case is about predicting whether a person's income is above \$50,000 a year based on their personal information. This system was biased against \textit{non-white women}. Hence, the sensitive variables are \textit{gender} and \textit{race}, the positive outcome is \textit{income higher than 50,000\$ a year}, and the unprivileged group is \textit{non-white women}.

\smallskip
\noindent \ding{228} \textbf{German Credit (GERMAN) \cite{hofmann2013uci} -}
This use case is about the adoption by a German bank of an ML system to predict the granting of credit. This system has been proven to be biased against women, \ie women have a lower probability of getting credit from the bank. In this case, the sensitive variable is \textit{gender}, and the unprivileged group is \textit{women}. The positive outcome is \textit{having a credit granted}. 

\smallskip
\noindent \ding{228} \textbf{Bank Marketing (BANK) \cite{moro2014data} -}
The Bank Marketing system was an ML system developed for a phone call company to predict whether the client would subscribe to a term deposit. This system was shown to be biased against people more than 25 years old. Hence, in this case, the sensitive variable is \textit{age}, and the unprivileged group is \textit{people with more than 25 years}. The positive outcome is \textit{will subscribe}. 


\smallskip
\noindent \ding{228} \textbf{Resyduo dataset (RESYDUO) \cite{di2023resyduo} -} 
This dataset comprises all the Arduino projects collected from the ProjectHub\footnote{\url{https://projecthub.arduino.cc/}} open-source repository. In particular, it includes 5,547 projects, 3,137 tags, 11,645 hardware components, and 1,802 libraries. It is worth mentioning that this data has been used to feed a collaborative filtering-based recommender system supporting Arduino project development \cite{di2023resyduo}.
Similar to the TPL dataset, the scope of the fairness analysis is to measure how popularity impacts the recommended items by considering two different sensitive variables for each project, \ie \texttt{views} and \texttt{respects} \revision{(\ie the number of appreciations from users)}. The former quantifies project popularity based on the number of users who view it. The latter represents explicit feedback on project quality. Consequently, fairness assessment can be conducted on two distinct disadvantaged groups, \ie \textit{low viewed} and \textit{low respected} projects.
\revision{Hence, in this use case, the sensitive variables are \textit{views} and \textit{respect}, and the positive outcome is \textit{recommendation}. The unprivileged groups are items with \textit{low views} and \textit{low respect}, respectively.}

\begin{table*}[ht!]
    \centering
    \caption{\revision{The examined use cases. Use cases adopted as running examples in the paper are highlighted in gray.}}
     \label{tab:useCases}
     \resizebox{\textwidth}{!}{
     \begin{tabular}{|l|l | l | l | c |} 
     \hline
         \textbf{Name} & \textbf{Domain} & \textbf{Sensitive attribute(s)} &  \textbf{Positive outcome} & \textbf{Existing approaches } \\  
         \hline
         COMPAS & Social  &  gender, race & Non-recidiv &  \makecell{ \cite{9793775},\cite{10.1145/3510003.3510202},\cite{9793943}, \cite{10.1145/3468264.3468537}, \cite{chen_maat_2022}, \cite{peng_fairmask_2023},
         \cite{10.1145/3550355.3552401}, \cite{10.1145/3468264.3468565}, \cite{10.1145/3510003.3510091}, \cite{chakraborty_fairway_2020}} \\ \hline

        ADULT & Social & race, gender & Income $>$ \$50.000  & 
         \makecell{\cite{10.1145/3377811.3380331}, \cite{10.1145/3238147.3238165}, \cite{10.1145/3510003.3510137}, \cite{aggarwal_testing_2021}, \cite{10.1145/3510003.3510202}, \cite{9793943} , \cite{chen_maat_2022}, \cite{peng_fairmask_2023}, \cite{10.1145/3468264.3468537}, \cite{10.1145/3550355.3552401}, \\ \cite{10.1145/3468264.3468565},\cite{10.1145/3510003.3510091},\cite{chakraborty_fairway_2020}, \cite{sharma_testing_2019},\cite{aggarwal_testing_2021},\cite{angell_themis_2018}} \\ \hline         
         
         GERMAN  & Financial & gender  & Credit granted &  \makecell{\cite{10.1145/3377811.3380331},\cite{aggarwal_testing_2021},\cite{10.1145/3510003.3510137}, \cite{10.1145/3510003.3510202},\cite{9793943}, \cite{10.1145/3468264.3468537}, \\ \cite{10.1145/3468264.3468537},\cite{10.1145/3550355.3552401}, \cite{chen_maat_2022}, \cite{peng_fairmask_2023}, \cite{10.1145/3468264.3468565}, \cite{10.1145/3510003.3510091}, \cite{chen_maat_2022}, \cite{chakraborty_fairway_2020},\cite{sharma_testing_2019}, \cite{angell_themis_2018}} \\ \hline         
      
        BANK & Financial & age & Client subscribed & \cite{10.1145/3510003.3510137}, \cite{10.1145/3510003.3510202},\cite{9793943}, \cite{chen_maat_2022}, \cite{peng_fairmask_2023}, \cite{10.1145/3468264.3468537}, \cite{10.1145/3377811.3380331}, \cite{10.1145/3510003.3510091}  \\ \hline
         RESYDUO & IoT & view, respect & Item recommended & \cite{di2023resyduo}\\ \hline
        \rowcolor{gray!25} \textbf{UNIVERSITY} & Education & gender & Positive admission & \cite{austin_will_2016}\\ \hline
        \rowcolor{gray!25} \textbf{TPL} & RSSE & frequency & Library recommended  & \cite{nguyen2023dealing}\\ \hline

     \end{tabular}
     }
 \end{table*}  

\subsection{Answer to RQ$_1$} \label{sec:rq1}

To answer RQ$_1$, we evaluate the existing approaches in terms of the elicited features introduced in Section \ref{sec:slr} \revision{(\ie \textbf{F1 - Bias definition}, \textbf{F2 - Domain bias definition}, \textbf{F3 - Custom metric definition}, \textbf{F4 - Metric composition}, \textbf{F5 - Automated fairness assessment}, \textbf{F6 - Tool availability})}. Furthermore, we discuss our approach by highlighting the contribution compared to the examined works shown in Table \ref{tab:toolComparison}.


\smallskip
\noindent
\textbf{F1 support  - Bias definition} Based on our investigations, it is evident that only five of the analyzed approaches can address both individual \circled{I} and group \circledb{G} fairness \rev{as shown in Table \ref{tab:toolComparison}}. These approaches are Fair-SMOTE, FairKit-learn, Astraea, FairML, and MANILA. In contrast, the other tools are designed to focus exclusively on either individual or group fairness. Notably, \tool stands out by offering modeling constructs that comprehensively cover both individual and group bias definitions. This flexibility is achieved by drawing upon essential concepts extracted from surveys and empirical studies \cite{mehrabi_survey_2021,caton_fairness_2020,wang_survey_2023}.  
\smallskip

\noindent
\textbf{F2 support - Domain bias definition} 
While tools like Fair-SMOTE, Fairkit-learn, FairML, or MANI\-LA provide methods to assess several canonical individual and group bias definitions automatically,
only ASTRAEA offers a dedicated grammar that allows the user to define any biases besides the traditional ones. \tool goes a step forward compared to ASTRAEA by providing a tailored metamodel conceived to define fairness at two different levels of abstractions, \ie domain-level and dataset-level (\eg refer to the evaluation of the RESYDUO use case in Section \ref{sec:rq2}, where, starting from the same domain, two different fairness analyses are depicted selecting two different sensitive attributes of the dataset).

\smallskip
\noindent
\textbf{F3 support -  Custom metric definition} 
It is worth noting that only two approaches within our analysis allow for the customization of metric definitions: Themis and Fairway. \tool provides the \texttt{Function} metaclass (see Section \ref{sec:metamodel}) to specify a dedicated operation on sensitive variables defined in the specification phase. Additionally, the metamodel incorporates a selection of significant metrics that have already been established in the literature, serving as valuable tools for assessing fairness in various contexts beyond the original ones. In essence, \tool can be used to evaluate the statistical parity of two sensitive groups that are not strictly bounded by the social domain. Furthermore, it empowers users to define novel bias metrics (\eg \textit{coverage} \cite{NGUYEN2019110460}) necessary for emerging domains, such as recommender systems, thereby adapting to evolving research needs and applications.

\smallskip
\noindent
\textbf{F4 support - Metric composition} Among the examined strategies, only MANILA supports this feature by modeling each metric as a feature and allowing their compositions by means of aggregation functions. Meanwhile, \tool relies on the metamodel to compose the defined metrics, thus pursuing the generalizability of the whole process in terms of entities and their combinations. 

\smallskip
\noindent
\textbf{F5 support - Automated fairness assessment} Although all the investigated tools offer automatic fairness assessment, these approaches are generally confined to established use cases and state-of-the-art metrics. In contrast, \tool is extendible and very flexible since it allows the conceptualization of fairness in domains known in the literature and domains not yet covered (\ie recommender systems and novel use cases). Furthermore, it allows for the creation of novel fairness metrics without sacrificing automation, making it a powerful and adaptable tool for addressing the evolving landscape of fairness assessment.

\smallskip
\noindent
\textbf{F6 support - Tool availability} Regarding the tool availability, it is important to note that all the scrutinized approaches offer a replication package, with the exceptions being TILE, AITEST, and ASTRAEA. The majority of these approaches utilize Python libraries and frameworks, primarily due to the fact that the tested models are machine learning-based. Furthermore, both FairML and MANILA go a step further by offering code generation functionalities that automate the deployment and testing of the system as defined during the setup phase. \tool adopts a comparable approach by utilizing dedicated Acceleo templates, which are fed with models adhering to the specified metamodel.

\begin{shadedbox}
\textbf{Answer to RQ$_1$:} Although offering a good degree of automation, existing approaches lack in supporting the customization of bias and fairness definitions. \tool fills this gap by covering all the elicited features for bias definition, fairness analysis specification, analysis implementation and fairness assessment.   
\end{shadedbox}

\subsection{Answer to RQ$_2$}\label{sec:rq2}

To address RQ$_2$, we have selected five distinct use cases from those previously discussed in Section \ref{sec:scenarios}, encompassing various application domains, including social, financial, education, recommender systems in software engineering (RSSE), and the Internet of Things (IoT). The details of the five use cases, implemented using \tool, are provided in Table \ref{tab:results}. Specifically, for each use case, we specify the chosen metric for assessment, the number of sensitive variables considered, and the \tool outcome. 

\revision{Note how two of these use cases (i.e., \textit{University} and \textit{TPL}) have already been implemented throughout the paper to show the main capabilities of \tool. \rev{In this Section, we report only the outcomes of the fairness assessment process, \ie \textit{Fair} and \textit{Biased} for the University and the TPL use cases, respectively.}}
\begin{table*}[ht!]
    \centering
    \caption{\rev{\tool implementation of the use cases. Use cases adopted as examples throughout the paper are highlighted.}}
    \label{tab:results}
    \resizebox{\textwidth}{!}{
    \begin{tabular}{|l|l|l|l|r|c|}
    \hline
     \textbf{Use case} & \textbf{Domain} & \textbf{Metric} & \makecell[c]{\textbf{Number of}\\\textbf{sensitive vars}} & \makecell[c]{\textbf{Outcome}\\ \textbf{(Expected)}} & \makecell[cc]{\textbf{\tool assessment}\\ \textbf{results}}  \\ 
     \hline
      COMPAS  & Social & \revision{Eq. Odds} & 2 \texttt{(sex,race)} & 0.3 ($\le |0.2|$) & \cellcolor{red!25}Biased   \\
      \hline
      \makecell[lc]{GERMAN\\(BIASED)}  & \multirow{2}{*}{Financial} & \multirow{2}{*}{EO} & \multirow{2}{*}{1 \texttt{(sex)}} & -0.25 ($\le |0.2|$) & \cellcolor{red!25}Biased   \\
      \cline{1-1} \cline{5-6}
      \makecell[lc]{GERMAN\\(DEBIASED)} &  & & & -0.05 ($\le |0.2|$) & \cellcolor{green!25}Fair  \\
      \hline
      \multirow{2}{*}{RESYDUO} & \multirow{2}{*}{IoT} & \multirow{2}{*}{GEI} & 1 (\texttt{views}) & 0.31 ($\ge |0.8|$) & \cellcolor{red!25}Biased    \\
      \cline{4-6} & &  & 1 \texttt{(respects)} & 0.28 ($\ge |0.8|$) & \cellcolor{red!25}Biased  \\
      \hline
      \rowcolor{gray!25}\textbf{UNIVERSITY} & Education & SP & 1 (\texttt{sex}) & -0.15 ($\le |0.2|$) &\cellcolor{green!25}Fair   \\
      \hline
      \rowcolor{gray!25} \textbf{TPL} & RSSE & COV  & 1 \texttt{(frequency)}& 0.29 ($= |1.2|$)  & \cellcolor{red!25}Biased    \\
      \hline
    \end{tabular}}
\end{table*}
Moreover, for the GERMAN use case, we conduct two separate analyses: the first utilizing the original biased dataset and the second involving the same dataset after applying a preprocessing algorithm designed to mitigate bias (specifically, the Debiaser for Multiple Variables \cite{daloisio_debiaser_2023}). This analysis scenario exemplifies a typical scenario for \tool, where a user initially assesses the fairness of the original dataset and subsequently verifies if the bias has been reduced after employing a debiasing method.\footnote{It is important to clarify that the mitigation of bias is beyond the scope of this approach, as \tool primarily focuses on designing and implementing the fairness assessment workflow, as outlined in Section \ref{sec:motivation}}
Finally, for the RESYDUO use case, we perform two distinct analyses, one considering the \textit{views} sensitive variable and the other focusing on the \textit{respects} sensitive variable. This approach showcases \tool's versatility and ability to handle different sensitive variables, further highlighting its capabilities.

\revision{The implemented models and generated code for each use case are reported in our replication package \cite{Daloisio_MODNESS_2024}.}

\begin{lstlisting}[caption=\revision{Bias definition for the COMPAS use case.},label=fig:compas-bias,style=DSLStyle]
GroupBias "compas"{
	definition: {
		domain: "justice";
		source:HUMAN_DISCRIMINATION;
		sensitiveVariables: {
			SensitiveVariable{
				name: "gender";
				values: "male","female";
			},
			SensitiveVariable{
				name: "race";
				values: "white","non-white";
			}
		};
		positiveOutcome: "Non Recidiv";
		unprivilegedGroup: {
			SensitiveGroup{
				name: "non-white men";
				sensitiveValue: "race.non-white", 
								"gender.male";
			};
		};
		privilegedGroup: {
			SensitiveGroup{
				name: "white women";
				sensitiveValue: "race.white",
								"gender.female";
			};
		};
    };
\end{lstlisting}

\begin{lstlisting}[style=DSLStyle,label=fig:compas-analysis,caption=\revision{Excerpt analysis definition for the COMPAS use case.}]
Dataset {
    id: 'compas';
    groundTruthLabelName: 'two-year-recid';
    predictedLabelName: 'prediction';
    filePath: 'compas.csv';
    positiveOutcome: {
        id: "non-recidiv";
        mappingOutcome: "Non Recidiv";
        value: { operator: EQUAL; value: 0.0; };
    };
    datasetSensitiveVariable: {
        DatasetSensitiveVariable{
            name: "sex";
            mappingSensitiveVariable: gender;
            values: {
                SensitiveVariableValue{
                    id: "female";
                    mappingValue: "gender.female";
                    value: { operator: EQUAL; value: 0.0; };
                },
                SensitiveVariableValue{
                    id: "male";						
                    mappingValue: "gender.male";
                    value: { operator: EQUAL; value: 1.0; };
                }
            }
        },
        DatasetSensitiveVariable{
            name: "race";
            mappingSensitiveVariable: race;
            values: {
                SensitiveVariableValue{
                    id: "white";
                    mappingValue: "race.white";
                    value: { operator: EQUAL; value: 1.0; };
                },
                SensitiveVariableValue{
                    id: "non-white";
                    mappingValue: "race.non-white";
                    value:{ operator: EQUAL; value: 0.0; };
                }					
            }
        }
    };	
};
datasetUnprivilegedGroup: {
        id: "non-white-men";
        mappingGroup: "non-white men";
        sensitiveVariables: ("compas.sex.female", 
        "compas.sex.male");
};
datasetPrivilegedGroup: {
    id: "white-women";
    mappingGroup: "white women";
    sensitiveVariables: ("compas.sex.male",
     "compas.race.white");	
};
\end{lstlisting}

\paragraph{The COMPAS use case.} Concerning the Social domain, we replicated the COMPAS use case. \revision{Listings \ref{fig:compas-bias} and \ref{fig:compas-analysis} shows the bias definition and an excerpt of the fairness analysis definition, highlighting how multiple sensitive variables and intersectional sensitive groups (\ie sensitive groups identified by more than one sensitive variable \cite{chen_fairness_2023}) can be defined in \tool.}
As previously described, this use case is about discrimination of \textit{non-white men} in the prediction of \textit{recidivism}. Hence, we modelled the bias definition in \tool, specifying \textit{gender} and \textit{race} as sensitive variables, \textit{non-recidiv} as the positive outcome, \textit{non-white men} as the unprivileged group, and \textit{white women} as the privileged group \revision{(see Listing \ref{fig:compas-bias})}.

Next, we defined our fairness analysis by specifying the dataset containing all the related information \revision{(see Listing \ref{fig:compas-analysis})}. In particular, we specified in the attributes of the \texttt{Dataset} class that the ground truth labels are encoded in the \texttt{two\_years\_recid} column. In contrast, the model predictions are encoded in the \texttt{prediction} column. Then, we modelled that the positive outcome equals \texttt{1.0}. Finally, we specified that the sensitive variables are encoded in the \texttt{race} and \texttt{sex}\footnote{We refer to the original column names of the dataset reported in \cite{angwin_machine_2016}} columns where \textit{non-white women} have a value of \texttt{1} for both columns. After defining the dataset, we specified the metrics for analysis. For this use case, we adopted the \revision{\textit{Equalized Odds (Eq. Odds)}\cite{hardt_equality_2016}} fairness definition, which is included in the \texttt{ExistingFairnessMetric} class. Finally, we specified that this metric should be equal to 0 to have fairness, with a tolerance value of 0.2. 

From such a model, \tool generates the code \revision{implementing the analysis. The code follows the same structure of Listing \ref{lst:university} and is reported in our replication package.}

\paragraph{The GERMAN use case.} Concerning the Financial domain, we implemented the GERMAN use case, which is about the discrimination of women in credit granting.
Similarly to the COMPAS use case, we first specified in the bias definition the sensitive variable (\textit{gender}), the positive outcome (\textit{credit grant}) and the privileged (\textit{men}) and unprivileged (\textit{women}) groups. Next, we specified two different fairness analyses, one involving the original biased dataset and another involving the debiased one. In both analyses, we modeled that the \texttt{sex} column encodes the \textit{gender} sensitive variable where \textit{women} have a value equal to \texttt{1}. In contrast, the positive outcome is encoded in the \texttt{credit} column with a value equal to \texttt{1}. In both analyses, we selected the \textit{Equal Opportunity (EO)} \cite{hardt_equality_2016} fairness definitions, specifying a threshold of 0 and a tolerance value of 0.2. \revision{The generated code follows the same structure of Listing \ref{lst:university} and is reported in the replication package as well as the \tool implementation for this use case.} 

\paragraph{The RESYDUO use case.} Finally, for the IoT domain, we implemented the RESYDUO use case, which is about popularity bias in recommending software and hardware Arduino components \cite{di2023resyduo}. In particular, the system can retrieve libraries in the long tail and expose them to projects. This increases the possibility of coming across serendipitous libraries \cite{10.1145/1864708.1864761}, e.g., those that are seen by chance but turn out to be useful for the project under development. For example, there could be a recent library, yet to be widely used, that can better interface with new hardware or achieve faster performance than popular ones. Therefore, our analysis is enough to consider only views and respect, as the other features do not affect the popularity bias. This is confirmed by our previous work on generic TPL recommendations \cite{nguyen2023dealing}.

In the bias definition, we specified \textit{views} and \textit{respect} as sensitive variables and \textit{high ranking} as the positive outcome. Next, we defined two sensitive groups: one identified by the \textit{views} sensitive variable (\ie the privileged group is \textit{high-viewed} items, while the unprivileged group is \textit{low-viewed} items), and one identified by the \textit{respect} sensitive variable (\ie the privileged group is \textit{high-respected} items, while the unprivileged group is \textit{low-respected} items). Further, we defined two different fairness analyses. The first one aims at assessing the amount of popularity bias with respect to the number of \textit{views}, while the second aims at assessing the popularity bias with respect to the level of \textit{respects}. In both analyses, we specified that the predicted rank is encoded in the \texttt{tot\_recommendations} column and that we consider a rank \textit{high} if it is greater than the 80\% of the predicted ranks (like done for the TPL use case, see Section \ref{sec:proposedApproach}). Next, in the first analysis, we specified that the number of views is encoded in the \texttt{views} column and that an item is \textit{highly viewed} if its number of views is higher than 80\% of the other items. Instead, in the second analysis, we specified that the level of respect is encoded in the \texttt{respects} column and that an item is \textit{highly respects} if it has a respect level higher than 80\% of the other items. In both analyses, we modelled a custom metric used in the RecSys literature named \textit{Generalized Cross Entropy (GEI)}\cite{deldjoo2019recommender}. This metric measures how the probability distribution of having an item of the privileged group recommended is different from the probability of having an item of the unprivileged group recommended. Following the metric definition, we specified that this metric should be equal to or greater than 0.8 to have fairness. As for the other cases, the generated code follows the same structure of Listing \ref{lst:university} and is reported in our replication package \revision{as well as the \tool DSL implementation}. 

Altogether, the performed fairness assessment confirms that the bias is correctly detected in all the considered use cases, meaning that \tool is capable of detecting the biases defined at the model level.

\begin{shadedbox}
\textbf{Answer to RQ$_2$:}
\tool has a level of \textit{expressiveness} and \textit{correctness} able to model and successfully evaluate use cases from various domains, including social, financial, RSSE, and IoT. Our experiments demonstrate the extensive range of \tool's ability to define bias and fairness in different domains and its capability to automatically generate the relative experiments and hence assess fairness in the considered use cases.
\end{shadedbox}

\subsection{Answer to RQ$_3$}\label{sec:rq3_ans}

To address RQ$_3$, we conducted a comparative analysis between \tool and two MDE-based baselines for fairness assessment, namely FairML \cite{10.1145/3550355.3552401} and MANILA \cite{manila2023}. 
\revision{Assessing the quality of MDE-based tools is daunting since they usually rely on tailored metamodels conceived for a specific application domain. Prior works have defined a set of quality metrics that investigate several aspects, such as expressiveness, completeness, or portability. Within the scope of our paper, we follow the criteria proposed in \cite{Bertoa2010QualityAF} to establish two dimensions for facilitating comparison: \textit{expressiveness} and \textit{automation}. We frame these aspects by referencing the set of features detailed in Section \ref{sec:feature}}
\revision{\begin{itemize}
    \item[--] \textbf{Expressiveness:} This dimension measures the extent to which the tool enables the modelling of bias definitions and relative fairness analysis (encompassing features \textbf{F2-F3-F4}).
    \item[--] \textbf{Automation:} This dimension evaluates the degree to which the tool streamlines the entire fairness assessment process (encompassing features \textbf{F5-F6} plus an additional feature describing the level of guidance for the user provided by the tool in the fairness analysis specification).
\end{itemize}}

\revision{These dimensions are assessed on a scale ranging from 1 to 3, based on the number of features provided by the tools for both \textbf{expressiveness} (\ie F2, F3 and F4) and \textbf{automation} (\ie F5, F6 and guidance in the specification).} \rev{Furthemore, we marked if the examined tool support (\tick) or not (\untick) the corresponding feature.}
%

We model with the two baselines the use cases used to answer RQ$_2$, and we use them to evaluate these features.
\revision{The comparison results, focusing on \textbf{Expressiveness} and \textbf{Automation}, are presented in Table \ref{tab:baselines}}.

\begin{table*}[ht!]
\centering
\caption{\revision{Baseline comparison. For each baseline, we evaluate the expressiveness and automation scores based on the features they provide.}}
\label{tab:baselines}
\resizebox{\textwidth}{!}{%
\begin{tabular}{|l|ccc||c||ccc||c|}
\hline
\multirow{2}{*}{} &
  \multicolumn{3}{c||}{\textbf{Expressiveness}} &
   &
  \multicolumn{3}{c||}{\textbf{Automation}} &
  \multicolumn{1}{c|}{\textbf{}} \\ \cline{2-9} 
 &
  \multicolumn{1}{c|}{\textbf{\begin{tabular}[c]{@{}c@{}}Abstract\\bias def.\end{tabular}}} &
  \multicolumn{1}{c|}{\textbf{\begin{tabular}[c]{@{}c@{}}Custom\\metric def.\end{tabular}}} &
  \textbf{Metric comp.} &
  \multicolumn{1}{c||}{\textbf{\begin{tabular}[c]{c}Expr.\\ Score\end{tabular}}} &
  \multicolumn{1}{l|}{\textbf{Tool available}} &
  \multicolumn{1}{c|}{\textbf{Code generation}} &
  \multicolumn{1}{c||}{\textbf{\begin{tabular}[c]{@{}c@{}}Spec. \\ guidance\end{tabular}}} &
  \multicolumn{1}{c|}{\begin{tabular}[c]{@{}c@{}}\textbf{Automation}\\\textbf{Score}\end{tabular}} \\ \hline
\textbf{FairML} &
  \multicolumn{1}{c|}{\untick} &
  \multicolumn{1}{c|}{\untick} &
   \untick &
   0 &
  \multicolumn{1}{c|}{\tick} &
  \multicolumn{1}{c|}{\tick} &
   \tick &
   3 \\ \hline
\textbf{MANILA} &
  \multicolumn{1}{c|}{\untick} &
  \multicolumn{1}{c|}{\untick} &
   \tick &
   1 &
  \multicolumn{1}{c|}{\tick} &
  \multicolumn{1}{c|}{\tick} &
  \tick &  
   3 \\ \hline
\textbf{MODNESS} &
  \multicolumn{1}{c|}{\tick} &
  \multicolumn{1}{c|}{\tick} &
  \tick &
   3 &
  \multicolumn{1}{c|}{\tick} &
  \multicolumn{1}{c|}{\tick} &
  \untick &
   2 \\ \hline
\end{tabular}%
}
\end{table*}

\textbf{Expressiveness} pertains to the extent to which the tools offer abstraction capabilities to model a variety of heterogeneous use cases. To assess this aspect, we implemented each of the use cases detailed in Section \ref{sec:rq2} with every baseline tool \revision{and assessed their ability to define high-level custom bias definitions and custom metric definitions and to compose different existing metrics. In the following, we describe in detail each tool and explain how they provide or not these features.}

\revision{Similarly to \tool, FairML relies on an MDE-based infrastructure to define fairness assessments using a dedicated DSL. In particular, the tool provides abstractions to specify standard metrics from the AIF360 library. However, the tool does not provide abstractions to define high-level custom bias definitions and custom metrics or to compose existing ones. Hence, FairML received \texttt{0} as score for \texttt{Expressiveness}.}

\revision{MANILA relies on the Extended Feature Model (ExtFM) formalism to model a fairness evaluation workflow as a Software Product Line. In particular, the ExtFM provides a set of features to compose a fairness evaluation, among which there is a set of fairness metrics to adopt. The tool employs the set of metrics from the AIF360 library. However, it does not provide features to compose custom metrics. Moreover, the tool does not provide features to define high-level custom bias definitions. Hence, neither custom bias definition nor custom metric features are supported. Instead, the tool provides a set of aggregation functions (like \textit{mean}, \textit{harmonic mean}, \textit{minimum}, or \textit{maximum}) to combine different metrics, providing the metric composition feature. Hence, MANILA received \texttt{1} as score for \textit{Expressiveness}.}

\revision{\tool instead has been developed to address the limitations of current baselines in defining and executing custom bias assessments. Hence, it provides abstractions to define high-level custom bias definitions and custom metrics and to compose existing ones. Moreover, like the two baselines, it provides abstractions to use existing metrics from the AIF360 library. Hence, \tool received \texttt{3} as expressiveness score.}

Regarding the degree of \textbf{automation}, both FairML and MANILA offer user guidance when defining fairness assessments. FairML, for instance, employs a decision tree to assist users in selecting the appropriate metric based on the analysis scope they intend to pursue. On the other hand, MANILA employs the Feature Model formalism to guide users in selecting a set of features that invariably results in a correct (i.e., executable) fairness experiment. 
As of the current development stage, \tool does not provide this level of user guidance, although it is worth noting that we have plans to expand the tool's capabilities in this regard (see Section \ref{sec:conclusion}). Consequently, we assigned a score of 3/3 for the automation level of FairML and MANILA, while \tool received a score of 2/3 for its current automation capabilities.

\begin{table}[ht!]
\centering
\caption{\revision{List of implemented use cases and assessment result.}}
\label{tab:comparison}
\resizebox{\linewidth}{!}{
\begin{tabular}{|l|r|r|r|}
\hline
 & \textbf{FairML} & \textbf{MANILA} & \textbf{\tool} \\ \hline
\textbf{COMPAS} & \cellcolor{red!25}0.28 &  \cellcolor{red!25}0.29 &  \cellcolor{red!25}0.3 \\ \hline
\textbf{GERMAN} &  \cellcolor{red!25}-0.2 &  \cellcolor{red!25}-0.23 &  \cellcolor{red!25}-0.25 \\ \hline
\makecell[l]{\textbf{GERMAN}\\\textbf{FAIR}} & \cellcolor{green!25}-0.05 & \cellcolor{green!25}-0.1 & \cellcolor{green!25}-0.05 \\ \hline
\textbf{UNIVERSITY} & \cellcolor{green!25}-0.15 & \cellcolor{green!25}-0.12 & \cellcolor{green!25}-0.15 \\ \hline
\textbf{TPL} & \untick & \untick & \cellcolor{red!25}0.29 \\ \hline
\textbf{RESYDUO} & \untick & \untick & \cellcolor{red!25}\makecell[r]{0.31 (\texttt{views})\\0.28 (\texttt{respect})} \\ \hline 
\end{tabular}
}
\end{table}

\revision{Finally, the list of assessment results for each use case implemented is presented in Table \ref{tab:comparison}. As can be seen, all the tools report comparable results for all the involved use cases. The small variability among the results can be explained by the different training-testing splits.} From this analysis, we have seen how all the selected baselines provide a high level of \textit{automation} in the definition and implementation of a fairness evaluation, \rev{\ie the outcome of the fairness assessment process is equal for all the three tools in the four notable use cases.} However, both baselines do not have a level of \textit{expressiveness} fairness analyses definition in terms of domains (\eg RSSE or IoT) and metrics (\eg \textit{coverage} or \textit{GEI}) \rev{as highlighted by the \untick \xspace in Table \ref{tab:comparison}}. \rev{Concerning the TPL use case, we obtained results close to the original ones, \ie the average coverage value for a single evaluation round is equal to 0.35\footnote{\url{https://github.com/MDEGroup/BiasInRSSE}}. Similarly to the previous use cases, the small variability may be explained by the different training-testing splits. and the size of the recommended items}





\begin{shadedbox}
    \textbf{Answer to RQ$_3$:} While all the examined baseline tools exhibit a high degree of automation throughout the fairness assessment process, both of them share a common limitation, \ie they lack the ability to express fairness in different domains. In contrast, \tool overcomes these limitations by offering a versatile framework for modelling high-level bias definitions and specifying and implementing custom fairness metrics tailored to specific application domains.
\end{shadedbox}


\section{Threats To Validity}
\label{sec:threats}
This section discusses possible threats that can hamper the results of the performed evaluation. 

\medskip
\noindent
\textit{Internal validity} concerns two aspects, \ie the light-weight survey and the conducted evaluation.

\revision{Concerning the lightweight survey presented in Section \ref{sec:slr}, considering only SE venues may lead to incorrect results, \eg excluding some relevant work from ML-related venues. To mitigate this issue, we carefully read and apply inclusion and exclusion criteria by focusing on papers that introduce a certain degree of automation in the fairness assessment process, focusing on metric composition and fairness definition aspects.}

\revision{Concerning the proposed approach,} it is possible that the metamodel we have created and the supporting tools are not extensive enough to cover all fairness assessment scenarios. However, we purposely considered different application domains, including the one related to the popularity bias of recommender systems in software engineering. Another potential threat of our study concerns the macro-sources of bias we cover. In particular, we address \textit{algorithmic bias} and \textit{unbalanced group bias} \cite{mehrabi_survey_2021,daloisio_debiaser_2023}, despite various other macro-sources of bias have been identified over the years, such as \textit{confounding variables bias} \cite{daloisio_debiaser_2023}. However, we believe that our approach covers most of the bias case studies documented in the literature, as they originate from macro-sources of bias that we address. In addition, our proposed metamodel is also designed to be extendable to model sources of bias that are not currently addressed. We acknowledge that the code generation may not be accurate due to some errors while running the Acceleo transformation, \rev{especially in the generation of custom fairness metrics. To mitigate this threat, we compared the results obtained for the TPL use case with the ones from the original paper, reporting a mild difference.} 


\medskip
\noindent
\emph{External validity} threats concern the generalizability of our approach. In this respect, the results obtained in this paper may be valid only for the considered datasets. To mitigate this threat, we diversified the datasets, which have been collected from different sources and domains. Furthermore, we demonstrated that the proposed approach could cover fairness conceptualization and assessment also in the software engineering domain by considering popularity bias in recommender systems. Another threat that may hamper the obtained results is the choice of the baselines, \ie we cannot conduct a quantitative comparison in terms of metrics. To mitigate this, we conduct a qualitative analysis by reimplementing the examined use cases using the selected approaches.

\section{Conclusion And Future Work}
\label{sec:conclusion}

In this paper, we introduced \tool, an innovative model-driven engineering (MDE) approach designed to facilitate the customization of fairness definitions. Leveraging a dedicated metamodel, \tool offers a two-tier abstraction framework capable of modeling and validating the entire fairness workflow, spanning from conceptualization to assessment. Additionally, \tool incorporates a generation module that automates the fairness assessment process.

Our research commenced with a comprehensive survey of fairness toolkits and strategies published in prominent software engineering venues. Surprisingly, none of the existing tools supported the critical features of custom fairness and metric definitions. To showcase the capabilities of \tool, we conducted a qualitative evaluation demonstrating its efficacy in supporting fairness assessments across various domains, including recommender systems and the Internet of Things, in addition to traditional applications. While we covered a limited number of use cases, our results suggest that our methodology exhibits a high degree of both \textit{expressiveness} and \textit{correctness}, effectively facilitating the entire fairness assessment workflow. Furthermore, we compared \tool with two MDE-based baseline approaches for fairness assessment and highlighted how \tool overcomes their limitations.

In future work, we aim to enhance the underlying code generator by integrating additional target frameworks and toolkits to support a broader range of programming languages and fairness metrics \rev{and support additional cutting-edge models, such as pre-trained models or large language models. In particular, fairness assessment can be applied to additional tasks, \eg text generation, natural language processing (NLP), or summarization, leveraging those advanced models.}. Furthermore, we intend to implement a recommendation system that guides users through the specification of bias definitions and fairness analyses by providing suggestions for sensitive variables and appropriate fairness metrics based on user inputs. 
Finally, we aspire to \rev{develop a web-based graphical interface to facilitate a user evaluation from both industry and academia involving real-world and national project scenarios, respectively. This aim to assess the usability of \tool and identify areas for further improvement.}

\begin{acknowledgements}

This work has been partially supported by the \EmeliotAck, by the \SoBigDataITAck, and by the \FringeAck.

\end{acknowledgements}


\bibliographystyle{spmpsci}
\bibliography{sn-bibliography}

\end{document}